\documentclass[reprint,preprintnumbers,nofootinbib,amsmath,amssymb,aps,floatfix]{revtex4-2}
\usepackage{txfonts}
\usepackage[T1]{fontenc}

\usepackage[dvipsnames]{xcolor}
\definecolor{red}{rgb}{0.9, 0,0}
\definecolor{cerulean}{rgb}{0., 0.42,0.9}
\definecolor{navy}{rgb}{0.05, 0.05,0.8}

\usepackage[colorlinks]{hyperref}
\hypersetup{
    colorlinks = true,
    citecolor  = red,
	linkcolor  = navy
}

\usepackage{slashed}
\usepackage{graphics}
\usepackage{amsmath}
\usepackage{amssymb}
\usepackage{latexsym}
\usepackage{mathtools}
\usepackage{dsfont}
\usepackage{hyperref}
\usepackage{amsfonts}
\usepackage{enumitem}
\usepackage{xstring} 
\usepackage{xspace} 
\usepackage[normalem]{ulem}
\usepackage{fontawesome}
\usepackage{subcaption}
\usepackage{braket}
\usepackage{bm}

\newcommand{\rms}{{\rm rms}}
\newcommand{\ideal}{{\rm ideal}}

\def\vec#1{\mathbf{#1}}
\newcommand{\unit}[1]{\vec{\hat{#1}}}
\newcommand{\un}[1]{\boldsymbol{\mathbf{#1}}}

\usepackage{subfiles} %

\begin{document}

\title{Astronomical Image Blurring from Transversely Correlated Quantum Gravity Fluctuations}

\author{Vincent S. H. Lee}
\email{szehiml@caltech.edu}
\affiliation{Walter Burke Institute for Theoretical Physics, California Institute of Technology, Pasadena, CA 91125, USA}
\author{Kathryn M. Zurek}
\email{kzurek@caltech.edu}
\affiliation{Walter Burke Institute for Theoretical Physics, California Institute of Technology, Pasadena, CA 91125, USA}
\author{Yanbei Chen}
\email{yanbei@caltech.edu}
\affiliation{Walter Burke Institute for Theoretical Physics, California Institute of Technology, Pasadena, CA 91125, USA}

\preprint{CALT-TH-2022-039}

\begin{abstract}
    Quantum fluctuations in spacetime can, in some cases, lead to distortion in astronomical images of faraway objects. In particular, a stochastic model of quantum gravity predicts an accumulated fluctuation in the path length $\Delta L$ with variance $\langle \Delta L^2\rangle\sim l_pL$ over a distance $L$, similar to a random walk, and assuming no spatial correlation above length $l_p$; it has been argued that such an effect is ruled out by observation of sharp images from distant stars. However, in other theories, such as the pixellon (modeled on the Verlinde-Zurek (VZ) effect), quantum fluctuations can still accumulate as in the random walk model while simultaneously having large distance correlations in the fluctuations. Using renormalization by analytic continuation, we derive the correlation transverse to the light propagation, and show that image distortion effects in the pixellon model are strongly suppressed in comparison to the random walk model, thus evading all existing and future constraints. We also find that the diffraction of light rays does not lead to qualitative changes in the blurring effect.
\end{abstract}

\maketitle
\newpage
\tableofcontents

\section{Introduction}
\label{sec:introduction}

Although a consistent quantum mechanical description of spacetime geometry is not yet fully developed, it can be anticipated that spacetime geometry will have quantum fluctuations, which will  
manifest as uncertainties in the macroscopic {distances} traveled by 
{light rays}.
Over a distance $L$, one naively expects that the length fluctuation is $\Delta L\sim l_p$~\cite{Misner_1973}, where $l_p=\sqrt{8\pi G}=10^{-34}$~m is the Planck length associated with the UV scale of gravity with $G$ being the gravitational constant. However, it is possible for length fluctuations to accumulate over an entire path of length $L$, like a random walk, 
leading to an overall uncertainty of~\cite{Amelino-Camelia:1994suz, Diosi:1989hy} 
\begin{equation}\label{eqn:variance}
	\langle \Delta L^2\rangle \sim l_pL \, .
\end{equation} 
More specifically, random walk can arise from a ``spacetime foam"~\cite{Wheeler_1955} model, in which the spacetime metric has independent, order-of-unity fluctuations at neighboring Planck-length intervals~\cite{Hawking:1979pi, Ashtekar_1992}.  During each Planck time $t_p=l_p/c$,
a photon
deviates from its classical 
path by a random step with zero average and  $ \sim l_p$  uncertainty~\cite{Wigner:1957ep, Salecker_1958, Kwon:2014yea}.  After $N = L/l_p$ steps, the variance of the total distance traveled by the light ray is then given by  $\Delta L  \sim \sqrt{N} l_p$, consistent with  Eq.~\eqref{eqn:variance}.  

It has been proposed that the phase front of starlight propagating in such a spacetime foam will be distorted and lead to blurring of telescope images~\cite{Lieu:2003ee, Ng:2003jk, Ragazzoni:2003tn, Christiansen:2005yg, Perlman:2011wv, Perlman:2014cwa, Hogan:2023rea, Steinbring:2023abf}.  However, diffraction of light is known to be able to restore the transverse coherence of the phase front~(see, e.g., Chapter 9 of Ref.~\cite{thorne2017modern}) and restore the quality of images for a large class of spacetime foam models~\cite{Chen:2014ueo}.

If, however, quantum gravity is holographic, then the number of spacetime degrees of freedom in a region of spacetime is set by the area of its boundary, instead of the volume of its bulk.  %
This implies spatial correlations in the quantum degrees-of-freedom of spacetime, manifesting as correlations between metric fluctuations at different spacetime locations.
In particular, Verlinde-Zurek (VZ) proposed how holographic theories could give rise to an accumulation of spacetime fluctuations consistent with Eq.~\ref{eqn:variance}~\cite{VZ1}.
In the VZ theory, quantum degrees of freedom on the boundary of a causal diamond drive quantum fluctuations in the spacetime geometry inside the diamond, which in turn leads to fluctuations in the size of the causal diamond~\cite{VZ2, Banks:2021jwj, Gukov:2022oed, VZ3}. In Ref.~\cite{VZ1}, VZ specifically considered the causal diamond generated by the union of the future and past domains of dependence, which in flat spacetime is simply a sphere.

For the radius $L$ of the spatial sphere along any arbitrary direction, while they found a magnitude of fluctuation $\Delta L$ along the lightcone directions consistent with Eq.~\ref{eqn:variance}, they also discovered unique long distance (large angle) angular correlations between radii along different angular directions such that most of the power of the quantum fluctuations lies in the low angular harmonic modes.   Decomposing the length fluctuation correlation between two points on a sphere with angular coordinates $\unit{\Omega}$ and $\unit{\Omega}'$ as a sum over spherical harmonics, the contribution from  each mode exhibits a $l^{-2}$ scaling for $l\gg 1$~\cite{VZ1}, {\it i.e.},
\begin{equation}\label{eqn:l_2_scaling}
	\langle\Delta L(\unit{\Omega})\Delta L(\unit{\Omega}')\rangle \sim \frac{l_pL}{4\pi}\sum_{lm}\frac{1}{l^2}Y^m_l(\vec{\Omega})Y^{m*}_l(\vec{\Omega}') \, .
\end{equation} 
This is consistent with the 't Hooft uncertainty relation~\cite{tHooft:1996rdg} that motivated the VZ proposal.  In particular, the modular energy fluctuations that generate a length fluctuation of Eq.~\ref{eqn:l_2_scaling}~\cite{VZ2} have been shown to have a physical origin in gravitational shock waves produced by vacuum energy fluctuations~\cite{VZ3,He:2023qha}.  %
These gravitational shock waves motivated the use of  't Hooft commutation relations as a way to quantize gravity at black hole horizons~\cite{Dray:1984ha, Dray:1985yt, tHooft:1996rdg, tHooft:2018fxg}, and has profound implications in various contexts of quantum gravity, including the AdS/CFT correspondence and black hole thermodynamics~\cite{Mertens:2017mtv, Lam:2018pvp}.

As a consequence of the different correlation structure, while both spacetime foam and the VZ effect give rise to length fluctuations 
parametrically
of the size of 
Eq.~\eqref{eqn:variance}, they have different properties and observable signatures. For instance, the random walk model due to spacetime foam predicts spatial correlations in all directions consistent with Brownian motion for paths that are separated by more than $l_p$ from each other, while the VZ effect has stronger correlations over large transverse distances up to $L$~\cite{VZ1} as predicted by Eq.~\ref{eqn:l_2_scaling}.

Since interferometers are extremely sensitive to tiny length fluctuations, they are the primary experimental candidates for detecting spacetime fluctuations~\cite{Amelino-Camelia:1994suz, Ng:1999hm,Hogan:2007pk,Kwon:2014yea}.  The correlation function~\eqref{eqn:l_2_scaling} for radii emanating from the same origin alone is not sufficient to provide predictions for all experiments that can be performed in a spacetime, including for interferometers.   For that general purpose, the {\it pixellon} model was proposed~\cite{Zurek:2020ukz}, in which a breathing-mode metric perturbation is prescribed to provide consistent results with~\eqref{eqn:l_2_scaling}.   The pixellon model has been applied to make predictions for current and future interferometer experiments~\cite{Zurek:2020ukz, Li:2022mvy}.
A summary of the VZ proposal is given in Ref.~\cite{Zurek:2022xzl}.

\begin{figure*}
	\includegraphics[width=1.\textwidth]{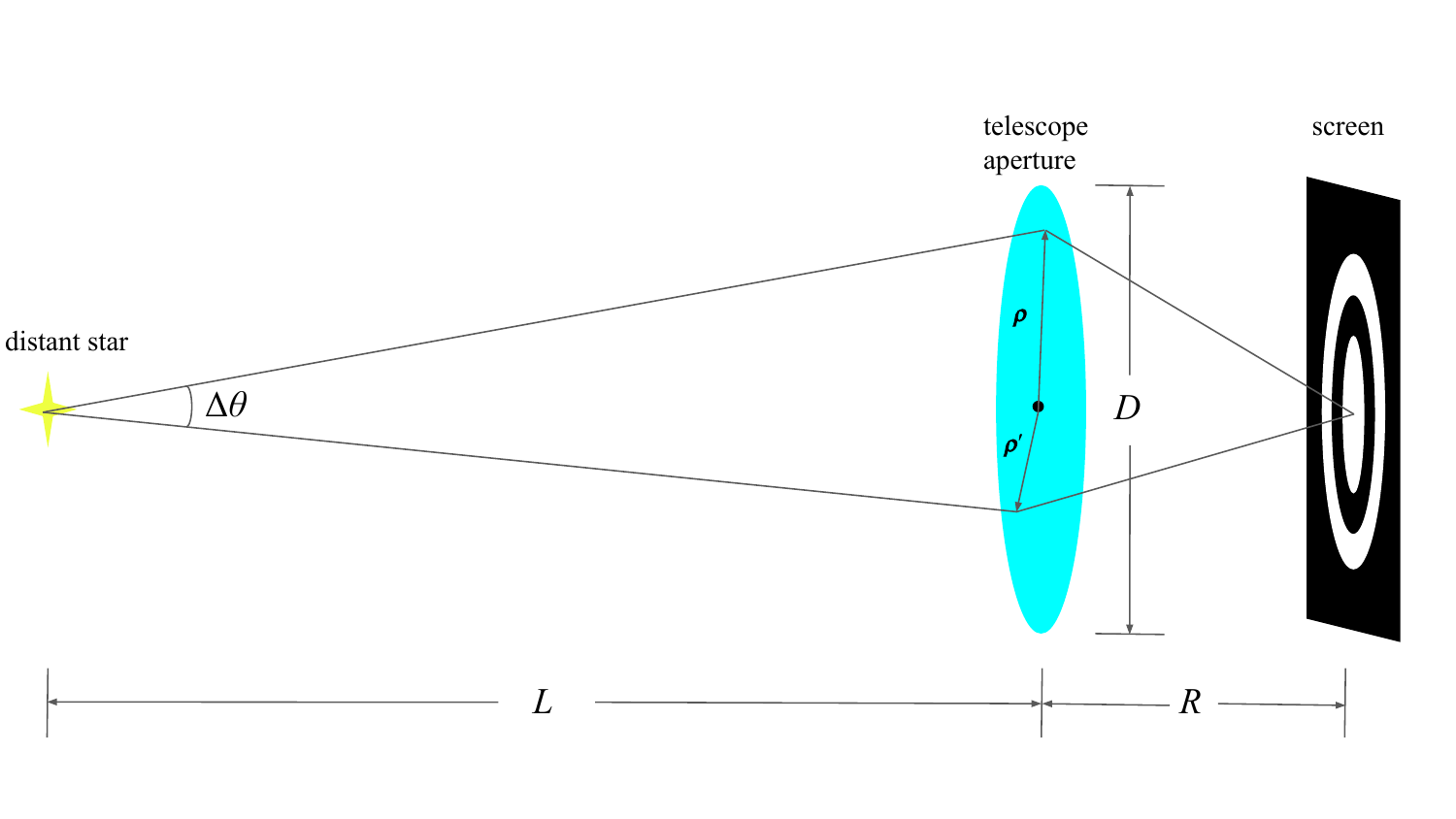} 
	\caption{Diagram showing light rays originating from a distant star passing through the telescope aperture and forming a sharp image on the screen. Here the star-aperture distance, aperture diameter, and aperture-screen distance are denoted by $L$, $D$, and $R$, respectively. Two distinct points on the aperture, denoted by two-dimensional vectors $\un{\rho}$ and $\un{\rho}'$, span an angle of $\Delta\theta$. Typically $L\ggg R\gg D$. The diagram is not drawn to scale.}\label{fig:setup}
\end{figure*}

In addition to interferometers,  it has been proposed that 
astronomical observations of distant stars can potentially constrain spacetime fluctuation thanks to the long propagation distance~\cite{Lieu:2003ee, Ng:2003jk, Ragazzoni:2003tn, Christiansen:2005yg, Perlman:2011wv, Perlman:2016xbc}.  
The fluctuating spacetime between an astronomical object and a telescope acts as a fuzzy lens, leading to degradation in image quality. Hence the observation of a diffraction-limited image places an upper limit on the magnitude of the fluctuations.  A diagram showing the propagation of light rays from a distant point source and the formation of its image in a telescope is given in Fig.~\ref{fig:setup}.

Assuming no spatial correlations %
between fluctuations at points separated by more than  $  l_p$, 
the upper limit $L$ for distance from a point source (a star, a galaxy, or a galaxy cluster) to the telescope, before the image of the star is blurred, can be naively estimated as  $\sqrt{l_pL}\lesssim \lambda_0$, where $\lambda_0$ is the optical wavelength.   For a cosmological distance of  $L\sim$ Gpc, we have $\sqrt{l_pL} \sim 5\times 10^{-5}$\,m, which is already far greater than the wavelength of visible light.

This estimate has been used to argue that the random walk model is thus completely ruled out by, for instance, existing data from the Hubble Space Telescope (HST)~\cite{Christiansen:2009bz, Beckwith:2006qi}. As an example, HST observed a star at a redshift of $z=6.2$~\cite{Welch_2022} with distance $L\sim 28$ Gly from Earth\footnote{Here we quoted the \textit{comoving} distance, which is argued in Ref.~\cite{Christiansen:2009bz} to be the correct distance measure for constraining spacetime foam as it measures the fabric of spacetime itself. However, as we will demonstrate in Sec.~\ref{sec:effcts_of_diffraction}, the limits from image blurring are satisfied by many orders of magnitude for quantum gravity models that we are interested in, and thus our conclusion does not depend on the choice of distance measure.}, corresponding to a length fluctuation of order $\sqrt{l_pL}\sim 0.1$ mm, far exceeding the optical wavelength $\lambda_0\sim1\,\mathrm{\mu m}$, and thus is expected to completely destroy the optical image of the star itself.  However, in the case of spacetime foam, the constraining power of astronomical image blurring has been questioned in several papers~\cite{Coule:2004qf, Dowker:2010pf}. In particular, Ref.~\cite{Chen:2014ueo} argues that wave diffraction of photons with sub-Planckian energy as they propagate through spacetime introduces an extra factor of $l_p/\lambda_0$ in the level of fluctuations, completely eliminating any hope in constraining spacetime foam with any physical system.

In this work, we use the pixellon model~\cite{Zurek:2020ukz, Li:2022mvy}  realization of the VZ effect to show that length fluctuations given in Eq.~\eqref{eqn:l_2_scaling} are not ruled out by the observations of astronomical images.  
The key idea is that while the phase shift relative to the classical photon path is large, the astrophysical image does not become blurred, since blurring depends not on the phase difference relative to the classical path, but rather on the phase difference between light rays hitting two typical points on the aperture, as illustrated in Fig.~\ref{fig:setup}. Since the telescope size $D$ is much smaller than the classical photon path length $L$, the phase difference  {\it across the aperture} due to quantum fluctuations parametrized by Eq.~\eqref{eqn:l_2_scaling}  is much smaller, and thus cannot sufficiently decohere the light to prevent the formation of a sharp image. Using this argument plus a Riemann zeta regularization, we will demonstrate that for $L >  D$,  the maximum distance of propagation before the image is destroyed is given by  
\begin{equation}\label{eqn:supprssed_limit}
	\sqrt{l_pL}\lesssim\frac{\lambda_0 L}{D} \, ,
\end{equation} 
which is easily satisfied by any realistic telescope.\footnote{Note that inequality~\eqref{eqn:supprssed_limit} can be formally violated when $L$ is chosen to be a very small distance.  However, that will violate the $L>D$ condition from which \eqref{eqn:supprssed_limit} was derived.}

It is natural to ask whether wave diffraction, as analyzed in Ref.~\cite{Chen:2014ueo} in the context of spacetime foam, affects our image blurring analysis when spatial correlations in the pixellon model are  taken into account. However, since in the pixellon model, the energy associated with length fluctuations over a distance $L$ mainly arises from scales $\sim 1/L$, which is much less than the optical frequency of a light ray, diffraction produces very little effect. This will be explicitly demonstrated using Huygens-Fresnel-Kirchhoff scalar diffraction theory~\cite{Jackson:1998nia}.

The rest of the paper is organized as follows. In Sec.~\ref{sec:transverse_correlation}, using the pixellon model described in Refs.~\cite{Zurek:2020ukz, Li:2022mvy}, we derive the transverse correlations of fluctuations (more specifically, two-point correlation functions) in optical path lengths connecting a distant point source and points on the aperture of the telescope (see Fig.~\ref{fig:setup}).  In Sec.~\ref{sec:blurring_effects}, we apply diffraction theory from the aperture of the telescope to the image plane and quantify the level of image degradation due to the pixellon-model-induced phase fluctuations.  We also present some numerical results on the blurring effects as observed on an opaque screen. In Sec.~\ref{sec:effcts_of_diffraction}, we consider the effects of diffraction between the point source and the telescope and show that they are subdominant for correlated fluctuations obtained in Sec.~\ref{sec:blurring_effects}. Finally, in Sec.~\ref{sec:conclusion}, we summarize our conclusions.

\section{Transverse Correlations of Length Fluctuations of Light Rays from a Distant Star}
\label{sec:transverse_correlation}

In this section, we derive the transverse correlations of the distances traveled by light rays emitted from a distant point source.  This source can either be a star or a galaxy, but for simplicity, we shall refer to it as a star.  The setup is shown in Fig.~\ref{fig:setup}. Assuming that the incident light is normal to the aperture plane, the blurring effect on the image provides a direct probe to the transverse correlation. 

We commence with a brief review of the pixellon model as described in Refs~\cite{Zurek:2020ukz, Li:2022mvy}, but we emphasize that our analysis holds for any quantum gravity model that produces angular two-point correlation functions with a spherical harmonic decomposition in the form of Eq.~\eqref{eqn:l_2_scaling}. The pixellon model describes the breathing mode of a spherical entangling surface bounding a causal diamond~\cite{Banks:2021jwj, Gukov:2022oed, Zhang:2023mkf}. Implications of the pixellon model on interferometer and astrophysical observables are studied in Refs~\cite{Li:2022mvy, Bub:2023bfi}. The metric fluctuation can be written as a scalar field in the radial component of the metric
\begin{align}\label{eqn:pixellon_metric}
	ds^2 = -dt^2 + [1-\phi(t,\vec{x})](dx^2+dy^2+dz^2) \, .
\end{align} 
Decomposing the scalar field into Fourier components
\begin{align}\label{eqn:pixellon_Fourier}
	\phi(t,\vec{x}) = l_p\int\frac{d^3\vec{p}}{(2\pi)^3}
	\frac{a_{\vec{p}}e^{-i\omega_{\vec{p}}t+i\vec{p}\cdot\vec{x}}+a^{\dagger}_{\vec{p}}e^{i\omega_{\vec{p}}t-i\vec{p}\cdot\vec{x}}}{\sqrt{2\omega_\mathbf{p}}}\, ,
\end{align} 
the creation and annihilation operators admit a two-point function of 
\begin{align}\label{eqn:two_point_operators}
	\langle a_{\vec{p}}a^{\dagger}_{\vec{p}'}\rangle = (2\pi)^3\left(1+\frac{a}{l_p\omega_{\vec{p}}}\right)\delta^3(\vec{p}-\vec{p}') \, ,
\end{align} 
and an on-shell dispersion relation, 
\begin{equation}
\label{eq:pix:dispersion} 
\omega_{\vec{p}}=c_s|\vec{p}|\,,\quad c_s\equiv 1/\sqrt{3}\,.
\end{equation}
Here $a$ is a constant characterizing the theoretical uncertainty of the model. Note that the second term in Eq.~\eqref{eqn:two_point_operators} corresponds to the pixellon occupation number and is much greater than unity. We refer readers to Refs.~\cite{Zurek:2020ukz, Li:2022mvy, Bub:2023bfi} for details of the pixellon model.

We now compute the two-point function of $\phi$ defined in Eq.~\eqref{eqn:pixellon_Fourier} by using the correlation function \eqref{eqn:two_point_operators} and the dispersion relation \eqref{eq:pix:dispersion}, obtaining
\begin{align}\label{eqn:pixellon_2}
	&\langle\phi(t,\vec{x})\phi(t',\vec{x}')\rangle \nonumber \\
	&= al_p\int\frac{d^3\vec{p}}{(2\pi)^3}\frac{1}{\omega_{\vec{p}}^2}\cos\left[\omega_{\vec{p}}(t-t')-\vec{p}\cdot(\vec{x}-\vec{x}')\right] \nonumber \\
	&=\frac{\alpha l_p}{32\pi^2}\frac{1}{|\vec{x}-\vec{x}'|}\Theta\left(|\vec{x}-\vec{x}'|-c_s|t-t'|\right)\, ,
\end{align} 
where $\Theta$ is the Heaviside step function, 
and we have redefined the normalization constant to be $\alpha\equiv (2\pi/c_s^2)a$. Note that the benchmark of the theory is $\alpha\sim\mathcal{O}(1)$. Alternatively, we can apply the plane wave decomposition formula $e^{i\vec{p}\cdot\vec{x}}=4\pi \sum_{lm}i^lj_l(|\vec{p}||\vec{x}|)Y^m_l(\unit{x})Y^{m*}_l(\unit{x}')$ to Eq.~\eqref{eqn:pixellon_2}, which gives
\begin{align}\label{eqn:pixellon_2_Fourier_lm}
	\langle\phi(t,\vec{x})&\phi(t',\vec{x}')\rangle = \frac{\alpha l_p}{ 2\pi^2c_s}\int_{-\infty}^{\infty}d\omega\,e^{-i\omega(t-t')} \nonumber \\
	&\sum_{l=0}^{\infty}\sum_{m=-l}^l j_l\left(\frac{\omega |\vec{x}|}{c_s}\right)j_l\left(\frac{\omega |\vec{x}'|}{c_s}\right)Y^{m*}_l(\unit{x})Y^{m}_l(\unit{x}') \, ,
\end{align} 
where we have integrated over the angular components of the momentum using the orthogonality relation of spherical harmonics, and extended to negative frequency using the identity $j_l(-x)=(-1)^lj_l(x)$. Remarkably, when $t=t'$, the integral over frequency in Eq.~\eqref{eqn:pixellon_2_Fourier_lm} can be performed exactly, leading to 
\begin{align}\label{eqn:pixellon_2_Fourier_lm_2}
	\langle\phi(t,\vec{x})\phi(t,\vec{x}')\rangle &= \frac{\alpha l_p}{2\pi}\sum_{l=0}^{\infty}\sum_{m=-l}^l\frac{1}{2l+1}Y^{m*}_l(\unit{x})Y^{m}_l(\unit{x}')\nonumber \\
	\times&\begin{dcases*}
		\frac{|\vec{x}|^l}{|\vec{x}'|^{l+1}} & if $|\vec{x}|\leq|\vec{x}'|$ \\
		\frac{|\vec{x}'|^l}{|\vec{x}|^{l+1}} & if $|\vec{x}|>|\vec{x}'|$ \\
	\end{dcases*} \, .
\end{align}

We now consider two photon paths originating from a distant star, extending over a distance $L$ to two distinct points on the aperture with angular coordinates $\unit{\Omega}$ and $\unit{\Omega}'$. Clearly, when the aperture is much smaller than the star-telescope distance, $D\ll L$, the curvature of a sphere with radius $L$ centered at the origin can be ignored when only the extent of the telescope is considered. The length shift of a photon path arriving at the telescope at time $t$ is then given by integrating the linearized metric perturbation in Eq.~\eqref{eqn:pixellon_metric} along the classical photon path, 
\begin{equation} 
	\Delta L(t,\unit{\Omega})=-\frac{1}{2}\int_0^Ldr\,\phi(t-L+r,r\unit{\Omega})\,.
\end{equation}
The two-point function of the accumulated length fluctuation is thus obtained by integrating Eq.~\eqref{eqn:pixellon_2_Fourier_lm_2} along radial coordinates $r$ and $r'$, leading to 
\begin{align}\label{eqn:pixellon_2_Fourier_lm_final}
	&\langle \Delta L(t,\unit{\Omega})\Delta L(t,\unit{\Omega}')\rangle  \nonumber \\
	&= \frac{1}{4}\int_0^Ldr\int_0^Ldr'\,\langle \phi(t-L+r, r\unit{\Omega})\phi(t-L+r', r'\unit{\Omega}')\rangle \nonumber \\
	&= \frac{1}{4}\int_0^Ldr\int_0^Ldr'\,\langle \phi(t, r\unit{\Omega})\phi(t, r'\unit{\Omega}')\rangle \nonumber \\
	&=\frac{\alpha l_p L}{4\pi}\sum_{l=0}^{\infty}\sum_{m=-l}^l\frac{1}{(l+1)(2l+1)}Y^{m*}_l(\unit{\Omega})Y^{m}_l(\unit{\Omega}') \, .
\end{align} 
Here the third line of Eq.~\eqref{eqn:pixellon_2_Fourier_lm_final} directly follows from the result in Eq.~\eqref{eqn:pixellon_2} and the triangle inequality, $|r\unit{\Omega}-r'\unit{\Omega}'|\geq|r-r'|$, alongside with $c_s<1$. The $l^{-2}$ scaling of the transverse correlation found here has been anticipated in Sec.~\ref{sec:introduction}, and is in accordance with previous works in Refs.~\cite{VZ1, VZ3, Li:2022mvy, Zhang:2023mkf}. Using the addition theorem of spherical harmonics, $\sum_{m=-l}^lY^{m*}_l(\unit{\Omega})Y^{m}_l(\unit{\Omega}')=\frac{2l+1}{4\pi}P_l(\unit{\Omega}\cdot\unit{\Omega}')$ where $P_l$ is the Legendre polynomial, we can rewrite Eq.~\eqref{eqn:pixellon_2_Fourier_lm_final} as
\begin{align}\label{eqn:legendre}
	\langle \Delta L(t,\unit{\Omega})\Delta L(t,\unit{\Omega}')\rangle  = \frac{\alpha l_p L}{16\pi^2}\sum_{l=0}^{\infty}\frac{1}{l+1}P_l(\cos\Delta\theta)\, ,
\end{align} 
where $\Delta\theta$ is the angular separation between $\unit{\Omega}$ and $\unit{\Omega}'$. Note that Eq.~\eqref{eqn:legendre} is independent of time. Additionally, we note that the summation over $l$ in Eq.~\eqref{eqn:legendre} can be analytically performed to yield  $\log(1+\csc(\Delta\theta/2))$, which diverges logarithmically as $\Delta\theta\to 0$ .

Since both points are confined on the telescope, their separation is bounded by $\Delta\theta\lesssim D/L\ll 1$, and hence one can expand the Legendre polynomials for small argument, $P_l(\cos\Delta\theta) = 1-\frac{1}{4}l(l+1)\Delta\theta^2+\cdots$, and Eq.~\eqref{eqn:legendre} becomes
\begin{align}\label{eqn:legendre_expanded}
	\langle \Delta L(\unit{\Omega})&\Delta L(\unit{\Omega}')\rangle = \nonumber \\
	&\frac{\alpha l_p L}{16\pi^2}\left(\sum_{l=0}^{\infty}\frac{1}{l+1}-\frac{1}{4}\Delta\theta^2\sum_{l=0}^{\infty}l+\cdots\right) \, .
\end{align}

The sums here are clearly divergent but can be properly regulated by some large $\Lambda$ serving as a physical cutoff on the $l$ modes. As we will demonstrate in Sec.~\ref{sec:blurring_effects}, the blurring effects from correlated fluctuations do not depend on the absolute phase of the light rays, but only the phase difference between two typical points on the aperture. The first sum in Eq.~\eqref{eqn:legendre_expanded} is independent of $\Delta\theta$, and hence is an unimportant term that will drop out of the observable ({\it i.e.} the path difference between two points) by introducing a local counterterm, similar to an extrinsic energy quantity. The second sum in Eq.~\eqref{eqn:legendre_expanded} depends on the system size, $\Delta\theta$, and has to be regulated accordingly. In particular, we do not expect our effective field theory description of the VZ effect, the pixellon model, to be valid at scales $l \gtrsim 1/\Delta \theta$.  Imposing a cutoff on $l\Delta\theta$ then leads to the following regulated sums
\begin{align}\label{eqn:regularization}
	&\langle \Delta L(\unit{\Omega})\Delta L(\unit{\Omega}')\rangle \nonumber \\
	&\to 
	\frac{\alpha l_p L}{16\pi^2}\left(-\frac{1}{4}\Delta\theta^2\sum_{l=0}^{\infty}le^{-l\Delta\theta/\Lambda}+\cdots\right) \nonumber \\
	&=\frac{\alpha l_p L}{16\pi^2}\left[-\frac{\Lambda^2}{4}+\frac{1}{48}\Delta\theta^2+\mathcal{O}\left(\frac{1}{\Lambda}\right)+\mathcal{O}(\Delta\theta^4)\right] \, .
\end{align} 
The divergent parts are now all independent of $\Delta\theta$, and hence have to drop out of the observable, which becomes
\begin{align}\label{eqn:path_difference_regulated}
	&\langle [\Delta L(\unit{\Omega})-\Delta L(\unit{\Omega}')]^2 \rangle \nonumber \\
	= &2[\Delta L(\unit{\Omega})\Delta L(\unit{\Omega})-\Delta L(\unit{\Omega})\Delta L(\unit{\Omega}')]\nonumber \\
	= &-\frac{\alpha l_p L}{384\pi^2}\Delta\theta^2 \, .
\end{align} 
Such a length fluctuation is strongly suppressed in astrophysical observations due to the extra factor $\Delta\theta^2\lesssim (D/L)^2$.

Interestingly, one can directly arrive at the result in Eq.~\eqref{eqn:path_difference_regulated} by applying the following identity to Eq.~\eqref{eqn:legendre_expanded}
\begin{equation}\label{eqn:identification}
	\sum_{l=0}^{\infty} l = 1+2+3+\cdots = -\frac{1}{12} \, .
\end{equation} 
This identity originates from the analytic continuation of the Riemann zeta function and is more familiar in the context of the Casimir force~\cite{Casimir_1948}.  The Riemann zeta function, $\zeta(s)\equiv \sum_{n=1}^{\infty}n^{-s}$, is only convergent for $s>1$, but can be analytically continued to include the entire complex plane except for a simple pole at $s=0$. The identification amounts to $\zeta(-1)=-1/12$ by analytical continuation. This technique is known as the Zeta function regularization, which has been widely used to regulate divergent series in QFT and quantum gravity (e.g. see Ref.~\cite{Hawking:1976ja} for application to the gravitational path integral by Hawking). The divergent series is assigned a physical and finite value by analytic continuation, which always yields a unique value regardless of the actual renormalization scheme.

Returning to the image blurring analysis, analogous to the Casimir effect, the precise UV physics in Eq.~\eqref{eqn:regularization} that cuts off $l\Delta\theta>\Lambda$ does not matter, since the quantum fluctuation that can be realistically measured by an experimental device is only sensitive to the IR physics. In addition, we note that while Eq.~\eqref{eqn:path_difference_regulated} appears to have a spurious minus sign, it will not affect the size of the image blurring effect, since the path difference only enters the observable as a phase of the light rays.

\section{Blurring Effects from Correlated Fluctuations}
\label{sec:blurring_effects}

In Sec.~\ref{sec:transverse_correlation}, we argued that the path difference between two light rays is the actual quantity responsible for blurring a sharp image from a distant star. We computed the variance of the path difference in Eq.~\eqref{eqn:path_difference_regulated} and found a severe suppression factor of $\sim\Delta\theta^2$. In this section, we justify this argument by an explicit computation of the blurring effect by the Huygens-Fresnel principle of wave optics to light propagation from the telescope's aperture to the image plane. We find that the corresponding upper limits placed on the size of quantum fluctuation are given by Eq.~\eqref{eqn:Strehl_limit}, and are satisfied by many orders of magnitude.

In a typical astronomical observation, a telescope with size $D$ is pointed towards the source such that the surface of the aperture is embedded in the transverse plane with respect to the propagation direction of the incoming photons, which points from the source to the telescope.  (See Fig.~\ref{fig:setup} for an illustration of the system.)

The source can then be treated as generating a perfect spherical wave.  In the absence of perturbations (either astrophysical or quantum-gravity-induced), the incoming wave at the aperture of the telescope can be well approximated as a plane wave.  
The intensity profile observed by the telescope is obtained by treating each point of the aperture as a spherical wavelet (Huygens-Fresnel principle) and computing their interference pattern by considering the path difference of each wavelet  
\begin{equation}\label{eqn:Huygens}
	I_{\ideal}(\un{\sigma}) \propto \left|\int_{\Omega_A}d^2\un{\rho}\, e^{-i\un{k}\cdot \un{\rho}}\right|^2 \, ,
\end{equation} 
where $I_{\ideal}(\un{\sigma})$ is the unperturbed image intensity at $\un{\sigma}$ on the screen, $\Omega_A$ is the domain of the aperture, and we define $\un{k}\equiv(2\pi/\lambda_0 R)\un{\sigma}$ with $R$ being the aperture-screen distance. Here we have assumed the far-field limit ($R\gg D$). Let $A=\pi(D/2)^2$ be the area of the aperture, the expression in Eq.~\eqref{eqn:Huygens} is more commonly written as the (squared) Fourier transform of the aperture function $w(\un{\rho})$, defined to be $1$ where the aperture is unblocked, and zero otherwise
\begin{align}\label{eqn:intensity_fourier_transform}
	I_{\ideal}(\un{\sigma}) &= \frac{I_{0}}{A^2} \left|\int_{-\infty}^{\infty}d^2\un{\rho}\,w(\un{\rho}) e^{-i\un{k}\cdot \un{\rho}}\right|^2 \nonumber \\ 
	&= \frac{I_{0}}{A^2} \left|\tilde{w}(\un{k})\right|^2 \, ,
\end{align} 
where $I_0$ is the peak intensity of $I_{\mathrm{ideal}}(\un{\sigma})$. We now apply this to a circular aperture with a diameter $D$. Let $\gamma\ll1$ be the angular position of $\un{\sigma}$ relative to the origin, which is placed at the center of the aperture. Then the intensity profile is given by integrating Eq.~\eqref{eqn:Huygens}
\begin{align}\label{eqn:airy_dsk}
	I_{\ideal}(\gamma) &= \frac{I_{0}}{\pi^2(D/2)^4}\left|\int_0^{D/2} \rho d\rho \int_0^{2\pi}d\varphi\, e^{-i\frac{2\pi}{\lambda_0}\rho\gamma\sin\varphi}\right|^2 \nonumber \\
	&= 4I_{0}\left[\frac{J_1(\pi\gamma D/\lambda_0)}{\pi\gamma D/\lambda_0}\right]^2 \, .
\end{align} 
The profile in Eq.~\eqref{eqn:airy_dsk} is known as the \textit{Airy disk}, where the first minimum is located at $\gamma\approx 1.22\,\lambda_0/D$. 

Image blurring happens when each wavelet from the aperture acquires a random fluctuating phase, $\Delta\Phi(\un{\rho}) = (2\pi/\lambda_0)\Delta L(\un{\rho})$. 
The resulting intensity now picks up a phase factor,
\begin{equation}\label{eqn:intensity_with_fluctuation}
	I(\un{\sigma}) = \frac{I_{0}}{A^2} \left|\int_{-\infty}^{\infty}d^2\un{\rho}\,w(\un{\rho}) e^{i\Delta\Phi(\un{\rho})}e^{-i\un{k}\cdot \un{\rho}}\right|^2 \, ,
\end{equation} 
with the expectation value
\begin{align}
	\langle I(\un{\sigma})\rangle = \frac{I_{0}}{A^2} \int_{-\infty}^{\infty}&d^2\un{\rho}\int_{-\infty}^{\infty}d^2\un{\rho}' \nonumber \\
	 &w(\un{\rho})w(\un{\rho}')e^{i(\Delta\Phi(\un{\rho})-\Delta\Phi(\un{\rho}'))}e^{-ik(\un{\rho}-\un{\rho}')} \, .
\end{align} 
Defining $\sqrt{\langle\Delta\Phi^2\rangle}=\Delta\Phi_{\rms}$, for uncorrelated noise, one expects significant image distortion to happen when $\Delta\Phi_{\rms}\gtrsim \pi$. This is usually quantified using the Strehl ratio, defined as the ratio between the perturbed and unperturbed peak intensity, $S = \frac{\langle I(\theta=0)\rangle}{I_0}$, which is given by~\cite{Sandler_1994}
\begin{align}\label{eqn:Strehl_evaluated}
	S = \frac{1}{A^2}\int_{-\infty}^{\infty}d^2\un{\rho}\int_{-\infty}^{\infty}d^2\un{\rho}' w(\un{\rho})w(\un{\rho}')e^{-\frac{1}{2}D_{\Phi}(\un{\rho},\un{\rho}')} \, ,
\end{align} 
where we defined the two-point function $D_{\Phi}(\un{\rho},\un{\rho}')=|\langle[\Delta\Phi(\un{\rho})-\Delta\Phi(\un{\rho}')]^2\rangle|$. It is clear that if $\Delta\Phi$ has no spatial correlation, then the Strehl ratio simply decays exponentially with the rms value of the phase~\cite{Perlman:2016xbc}, $S_{\mathrm{uncorrelated}} = e^{-\Delta \Phi_{\rms}^2}$. Requiring $S$ to be close to unity then implies the rough estimate $\Delta\Phi_{\rms}\lesssim 1$ and thus $\sqrt{l_pL}\lesssim \lambda_0$, ignoring $\mathcal{O}(1)$ factors. This is a substantial level of fluctuations that have been ruled out, as we have discussed in Sec.~\ref{sec:introduction}. 

If we now take transverse correlation into account, then Eq.~\eqref{eqn:path_difference_regulated} implies the phase difference variance to be
\begin{equation}\label{eqn:phase_variance}
	D_{\Phi}(\un{\rho},\un{\rho}') = 2\Phi_{\rms}^2\frac{|\un{\rho}-\un{\rho}'|^2}{L^2} \, ,
\end{equation} 
where $\Delta\Phi_{\rms}^2= (1/192)(\alpha l_p L/\lambda_0^2)$. Combined with Eq.~\eqref{eqn:Strehl_evaluated}, one finds
\begin{equation}\label{eqn:Strehl_correlated}
	S = \frac{1}{A^2}\int_{\Omega_A}d^2\un{\rho}\int_{\Omega_A}d^2\un{\rho}' e^{-\Delta\Phi_{\rms}^2|\un{\rho}-\un{\rho}'|^2/L^2} \, .
\end{equation} 
Because $|\un\rho| < D\ll L$, $D_{\Phi}(\un{\rho},\un{\rho}')$ in Eq.~\eqref{eqn:phase_variance} is suppressed by an additional factor of $(D/L)^2$, which means the level of variation between $\Delta \Phi(\un{\rho})$ across the aperture is a factor $D/L$ suppressed from $\Phi_{\rm rms}$.  This drives the Strehl ratio strongly towards unity. 
More specifically, expanding the exponential in Eq.~\eqref{eqn:Strehl_correlated} gives
\begin{align}\label{eqn:Strehl_final}
	S &= 1-\frac{1}{192}\frac{\alpha l_p }{\lambda_0^2 L}\frac{1}{A^2}\int_{\Omega_A}d^2\un{\rho}\int_{\Omega_A}d^2\un{\rho}' \, |\un{\rho}-\un{\rho}'|^2 \nonumber \\
	&= 1 - \frac{1}{768}\frac{\alpha l_p D^2}{\lambda_0^2 L} \, .
\end{align} 
The formation of a sharp image from a distant star indicates $S\approx 1$ in Eq.~\eqref{eqn:Strehl_final}, and thus places a limit
\begin{equation}\label{eqn:Strehl_limit}
	\alpha \lesssim 3\times 10^{50}\,\left(\frac{\lambda_0}{1\,\mathrm{\mu}\mathrm{m}}\right)^2\left(\frac{L}{1\,\mathrm{Gpc}}\right)\left(\frac{1\,\mathrm{m}}{D}\right)^2 \, ,
\end{equation} 
which is clearly satisfied by many orders of magnitude in any realistic system with $\alpha\sim\mathcal{O}(1)$.
\begin{figure*}
	\includegraphics[width=1.\textwidth]{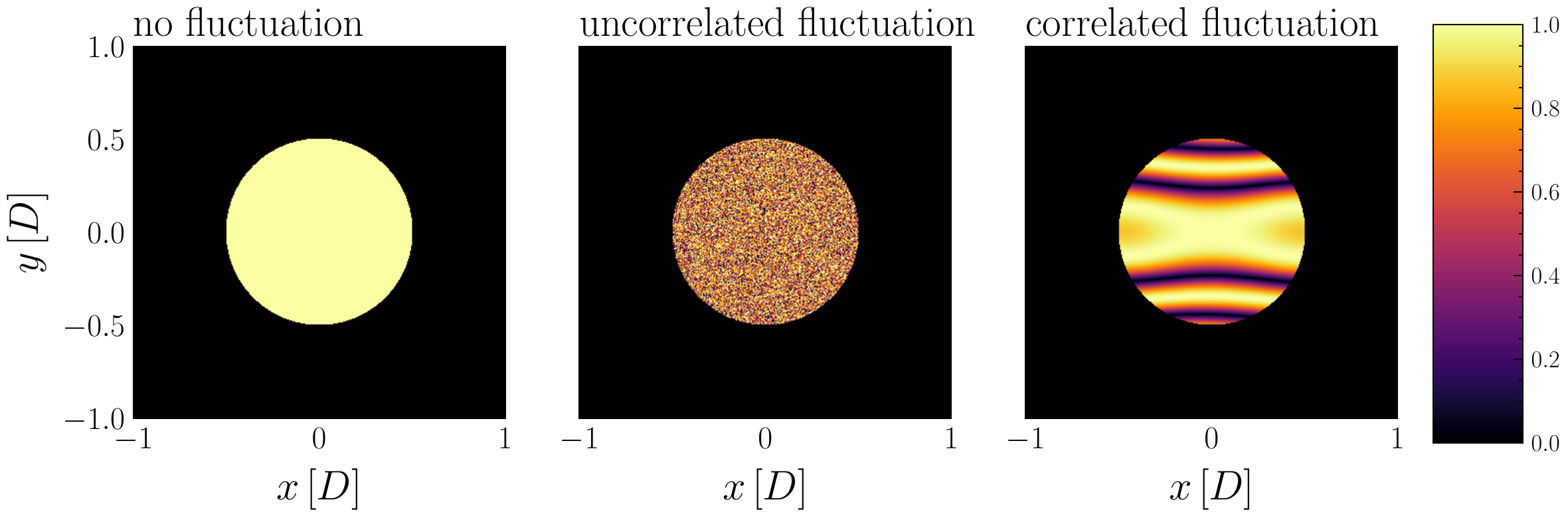} 
	\caption{Plot of a realization of $\cos\Delta\Phi(r)$ on the aperture, generated in accordance with Eq.~\eqref{eqn:phase_variance} assuming $\Delta\Phi_{\rms}=2\pi$. The rightmost panel assumes $L=D$.
	}\label{fig:aperture}
\end{figure*}
\begin{figure*}
	\includegraphics[width=1.\textwidth]{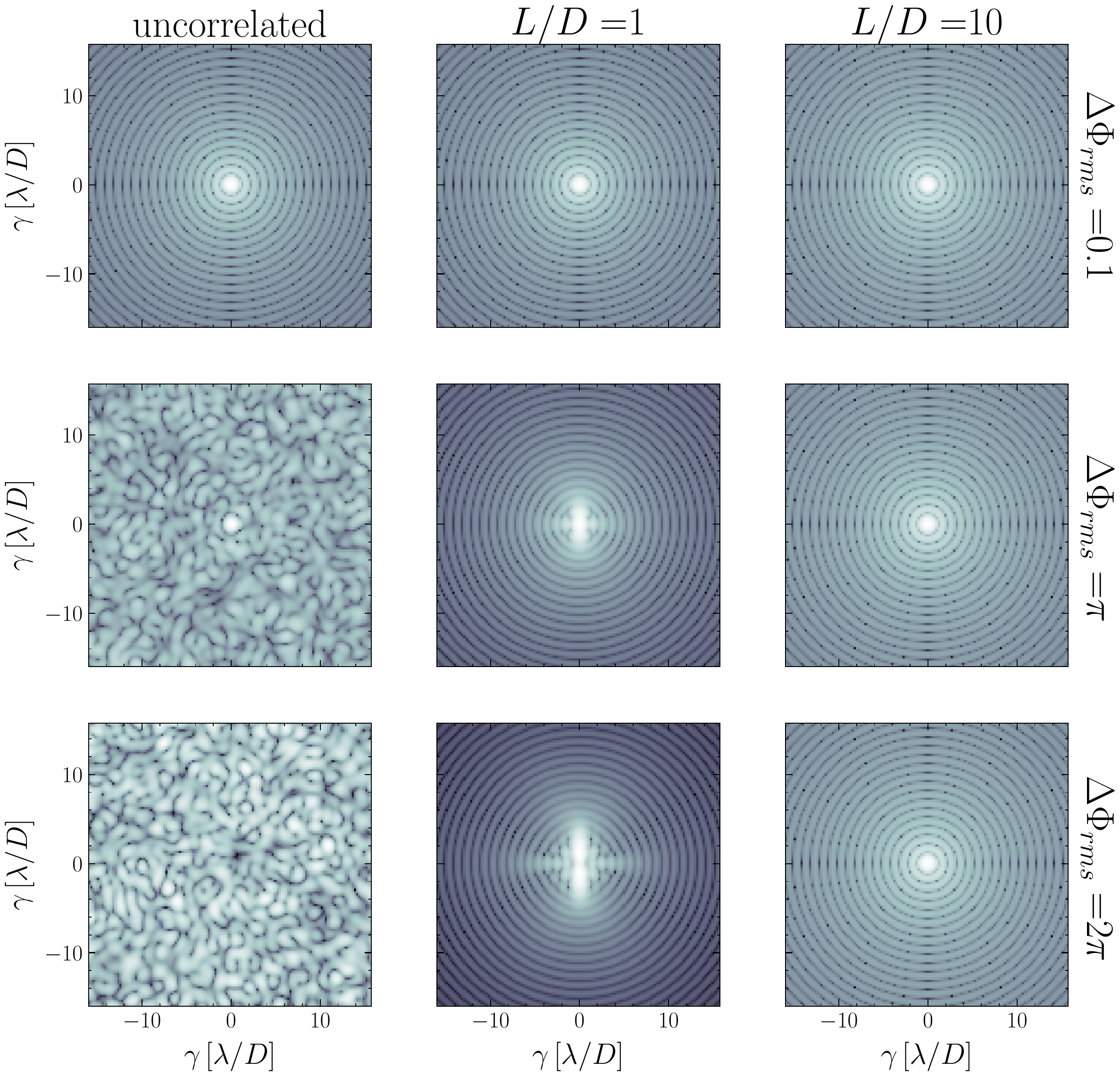} 
	\caption{Images from a point object for various values of $\Delta\Phi_{\rms}$ and correlation scale $L$, with correlation given by Eq.~\eqref{eqn:phase_variance}. The color represents $\log_{10}(I)$ with arbitrary normalization. The aperture is simulated with $1028\times1028$ pixels embedded in a $4096\times4096$ grid. The images are enlarged into the center $128\times 128$ pixels to better resolve the Airy disks. }\label{fig:image}
\end{figure*}

We perform a numerical simulation by generating random fields $\Delta\Phi(\un{\rho})$ across the aperture and producing the images as observed on the telescope's image plane by computing the Fourier transform in Eq.~\eqref{eqn:intensity_with_fluctuation}. We consider a circular aperture with a diameter of 1024 pixels, embedded in a $4096\times 4096$-pixel square. The phase fluctuation is assumed to be a random Gaussian field with $\Delta\Phi_{\rms}=0.1$, $\pi$ and $2\pi$. Correlated noise is generated to satisfy the variance in Eq.~\eqref{eqn:phase_variance}\footnote{This can be achieved numerically, for example, by proposing that $\langle \Delta\Phi(\un{\rho})\Delta\Phi(\un{\rho}')\rangle = \Phi_{\rms}^2e^{-|\un{\rho}-\un{\rho}'|^2/L^2}=(2\pi)^{-2}\int d^2\un{k}\,G(\un{k})e^{-i\un{k}\cdot(\un{\rho}-\un{\rho}')}$, where $G(\un{k})=(\pi\Phi_{\rms}^2L^2)e^{-L^2|\un{k}|^2/4}$. Correlated noise can then be numerically simulated by filtering a realization of \textit{uncorrelated} noise by the Green's function $G(\un{k}$).} assuming $L/D=1$ or 10.  We show the distribution of $\cos\Delta\Phi(\un{\rho})$ over the aperture in Fig.~\ref{fig:aperture} and plot the observed image in Fig.~\ref{fig:image}. When the noise is uncorrelated, the Airy disk patterns from the point source are destroyed when $\Delta\Phi_{\rms}\gtrsim\pi$, corresponding to a small Strehl ratio. However, once spatial correlation is introduced to the noise, the Airy disk patterns are restored, showing that correlated length fluctuations are much harder to constrain by studying image blurring effects.

\section{Effects of Diffraction}
\label{sec:effcts_of_diffraction}

In Sec.~\ref{sec:transverse_correlation}, we treated photons from a distant star as point particles (with zero wavelength) following null rays until they reached the telescope.  The sole contribution to the phase difference across the aperture is due to fluctuations in the distance covered by the rays. 
In other words, we have neglected the wave nature of the photons as they travel from the source to the telescope --- even though in Sec.~\ref{sec:blurring_effects} we have incorporated that wave nature as light propagates from the aperture to the screen.  In this section, we incorporate the wave nature of light rays more carefully by applying the Huygens-
Fresnel-Kirchhoff scalar diffraction theory~\cite{Jackson:1998nia} to light propagation from the source to the telescope, which has been argued in Ref.~\cite{Chen:2014ueo} to significantly modify the blurring effect in the context of spacetime foam. A recent work that investigates the effects of spacetime fluctuations on light rays, taking into account the wave nature of photons, is given in Ref.~\cite{Sharmila:2023ikb}.

\begin{figure*}
	\includegraphics[width=1.\textwidth]{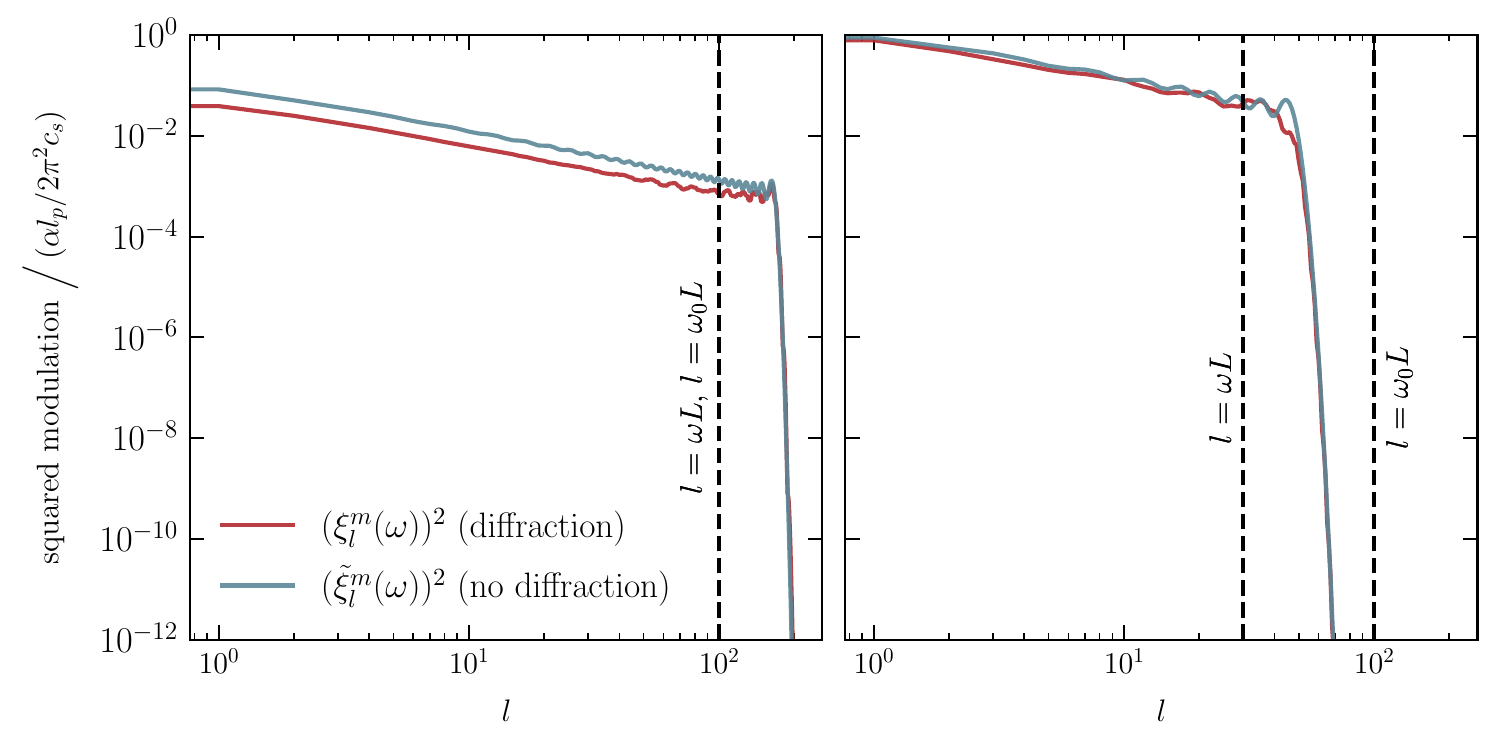} 
	\caption{Plots of the modulations on the aperture, $(\xi^m_l(\omega))^2$ (red, Eq.~\eqref{eqn:total_modulaion_3}) and $(\tilde{\xi}^m_l(\omega))^2$ (blue, Eq.~\eqref{eqn:no_diff_modulation}), as functions of $l$ with and without considering diffraction respectively. The curves are obtained by numerical integration. The left panel assumes $\omega_0 L=\omega L=100$ while the right panel assumes $\omega_0L=100$ and $\omega L=30$. The $y$ axes are normalized to $\alpha l_p/2\pi^2c_s$.}\label{fig:diffraction}
\end{figure*}

We compute and compare the size of phase modulations on the telescope aperture, as well as their transverse correlations, explicitly in Fig.~\ref{fig:diffraction}, and conclude that diffraction does not lead to qualitative changes in the observable if quantum gravity fluctuations have the high spatial correlation contained in the pixellon model. 

Consider a scalar wave $\Phi(t,\vec{x})$ with angular frequency $\omega_0$ propagating outwards in the radial direction with the metric in Eq.~\eqref{eqn:pixellon_metric}. The equation of motion of the scalar wave is given by $\frac{1}{\sqrt{-g}}\partial_\mu (g^{\mu\nu}\partial_\nu \Phi) =0$. The leading order contributions of the derivatives are the frequency of the scalar wave $\omega_0$, which allows us to take $\partial \phi \ll \partial \Phi$, leading to the familiar wave equation
\begin{equation}\label{eqn:wave_equation}
	\left[-(1-\phi(t,\vec{x}))\partial_t^2+\nabla^2\right]\Phi(t,\vec{x})=0 \, .
\end{equation} 
Note that one can also derive Eq.~\eqref{eqn:wave_equation} by setting $ds^2=0$ in Eq.~\eqref{eqn:pixellon_metric} and observing that the scalar wave travels with speed $dr/dt=(1-\phi)^{-1/2}$. Writing the wave as $\Phi(t,\vec{x})=[\Phi_0(\vec{x})+\psi(t,\vec{x})]e^{-i\omega_0 t}$, where $\Phi_0(\vec{x})=e^{i\omega_0 |\vec{x}|}/(4\pi |\vec{x}|)$ is an unperturbed spherical wave, and $\psi$ is the time-dependent scattered wave, the wave equation to the first order gives
\begin{equation}\label{eqn:first_order_wave_equation}
	(\nabla^2+\omega_0^2)\psi(t,\vec{x}) + (2i\omega_0\partial_t-\partial_t^2)\psi(t,\vec{x}) = \omega_0^2\phi(t,\vec{x})\Phi_0(\vec{x}) \, .
\end{equation} 
We can decompose the scattered wave and the pixellon field into Fourier and harmonic modes, {\it i.e.} $\psi(t,r\unit{\Omega}) = \int_{-\infty}^{\infty}d\omega\,e^{-i\omega t}\sum_{l=0}^{\infty}\sum_{m=-l}^l\psi^m_l(\omega,r)Y^m_l(\unit{\Omega})$ and $\phi(t,r\unit{\Omega}) = \int_{-\infty}^{\infty}d\omega\,e^{-i\omega t}\sum_{l=0}^{\infty}\sum_{m=-l}^l\phi^m_l(\omega,r)Y^m_l(\unit{\Omega})$. Then Eq.~\eqref{eqn:first_order_wave_equation} becomes
\begin{equation}\label{eqn:omega_wave_equation}
	\left[\nabla_r^2+(\omega_0+\omega)^2-\frac{l(l+1)}{r^2}\right]\psi^m_l(\omega,r) = \frac{\omega_0^2e^{i\omega_0 r}}{4\pi r}\phi^m_l(\omega,r) \, .
\end{equation} 
Imposing a regularity condition at the origin and  out-going wave boundary condition at infinity, we obtain a solution~\cite{Chen:2014ueo}:
\begin{align}\label{eqn:wave_equation_solution}
	\psi^m_l(\omega,L) = &\frac{\omega_0^2(\omega_0+\omega)h^{(1)}_l[(\omega_0+\omega)L]}{4\pi} \nonumber \\
	&\int_0^Ldr\left\{rj_l[(\omega_0+\omega)r]e^{i\omega_0r}\phi^m_l(\omega, r)\right\} \, .
\end{align} 
On the other hand, the variance of the pixellon modes has been derived in Eq.~\eqref{eqn:pixellon_2_Fourier_lm}
\begin{align}\label{eqn:pixellon_modes}
	\langle\phi^m_l(\omega,r)&\phi^{m'*}_{l'}(\omega',r')\rangle \nonumber \\
	 &= \frac{\alpha l_p}{2\pi^2c_s}j_l\left(\frac{\omega r}{c_s}\right)j_l\left(\frac{\omega' r'}{c_s}\right)\delta_{ll'}\delta_{mm'}\delta(\omega-\omega') \, .
\end{align} 
The modulation as measured at the telescope can be written as $\xi(t,\unit{\Omega})\equiv \psi(t,L\unit{\Omega})/\Phi_0(t,L\unit{\Omega})=4\pi L\psi(t,L\unit{\Omega})e^{-i\omega_0 L}$. Combining Eq.~\eqref{eqn:wave_equation_solution} and Eq.~\eqref{eqn:pixellon_modes}, the variance of the modulation is given by
\begin{align}\label{eqn:total_modulaion_2}
	&\langle \xi(t,\unit{\Omega})\xi(t,\unit{\Omega}')\rangle \nonumber \\
	&= \int_{-\infty}^{\infty}d\omega\sum_{l=0}^{\infty}\sum_{m=-l}^{l}{|\xi^{m}_l(\omega)|}^2Y^m_l(\unit{\Omega})Y^{m'}_l(\unit{\Omega}') \, ,
\end{align} 
where
\begin{align}\label{eqn:total_modulaion_3}
	 \left|\xi^{m}_l(\omega)\right|^2= &\left|\omega_0^2(\omega_0+\omega)Lh^{(1)}_l[(\omega+\omega_0)L]\right|^2\left(\frac{\alpha l_p}{2\pi^2 c_s}\right)\nonumber \\
	&\left|\int_0^Ldr\,re^{i\omega_0r}j_l[(\omega_0+\omega)r]j_l\left(\frac{\omega r}{c_s}\right)\right|^2 \, .
\end{align} 
Here $\xi_{l}^m$'s independence on $m$ results from the spherical symmetry of the two-point function.  While the integral over frequency in Eq.~\eqref{eqn:total_modulaion_3} is difficult to perform, we can estimate the variance in different regimes using asymptotic limits of Bessel functions. Realizing that each mode of the scattered wave $\psi^m_l(\omega)$ has a frequency of $\omega_0+\omega$, significant blurring effects of the image can only be induced by scattered waves with roughly the same frequency as the unperturbed wave, {\it i.e.} $\omega\lesssim \omega_0$. This leads to the following estimate for Eq.~\eqref{eqn:total_modulaion_3} 
\begin{align}\label{eqn:total_modulaion_4}
	&|\xi^{m}_l(\omega)|^2 \sim \left(\frac{\alpha l_p}{2\pi^2 c_s}\right)  \nonumber \\
	\times&\begin{dcases*}
	 	\left(\frac{\omega_0c_s}{2\omega}\right)^2\frac{\mathcal{O}(1)}{l} & if $l\ll (\omega L/c_s), \omega_0L$ \\
	 	\left(\frac{C_l}{2l+1}\right)^2\left(\frac{\omega L}{c_s}\right)^{2l}(\omega_0L)^4 & if $l\gg (\omega L/c_s),\,\omega_0 L$ \\ 
	\end{dcases*} \, ,
\end{align} 
where
\begin{align}\label{eqn:C_1l}
	C_l \equiv \frac{2^{l-1}l!}{(l+1)(2l+1)(2l)!} \ll 1 \, .
\end{align} 
Here $\mathcal{O}(1)$ in Eq.~\eqref{eqn:total_modulaion_4} denotes a number of order unity that has to be evaluated for each value of $l$. In Fig.~\ref{fig:diffraction} we plot $|\xi^{m}_l(\omega)^2|$ as a function of $l$ in blue for different choices of $\omega_0L$ and $\omega L$ (see caption), obtained by numerically integrating Eq.~\eqref{eqn:total_modulaion_3}. Estimates in Eqs.~\eqref{eqn:total_modulaion_4}-\eqref{eqn:C_1l} show that high $l$ modes with $l\gg \omega_0L$ are extremely suppressed by factors of $1/l!$ (more dramatic than an exponential suppression), which are negligible even when summed to $l\to\infty$. While we did not derive approximations in the intermediate $l$ regime where $l\ll \omega L/c_s$ but $l\gg \omega_0L$, it is clear from the right panel of Fig.~\ref{fig:diffraction} that these $l$ modes are also highly suppressed in this middle regime. Therefore, only the low $l$ modes satisfying $l\ll \omega L/c_s$ and $l\ll \omega_0L$ will contribute to image blurring.

We now compare these new results with those obtained in Sec.~\ref{sec:transverse_correlation}, where the effects of diffraction are neglected. In that approximate treatment, the phase modulation as measured at the aperture, denoted by $\tilde{\xi}(t,\unit{\Omega})$, was given by the photon path length fluctuation divided by the photon wavelength, {\it i.e.} $\tilde{\xi}(t,\unit{\Omega})=-(\omega_0/2)\int_0^Ldr\,\phi(t,r\unit{\Omega})$.  We can obtain the mode decomposition of $\xi$ by directly integrating Eq.~\eqref{eqn:pixellon_modes} over the radial distance, obtaining 
\begin{align}\label{eqn:no_diff_modulation}
	(\tilde{\xi}^{m}_l(\omega))^2 = \left(\frac{\alpha l_p}{2\pi^2 c_s}\right)\left(\frac{\omega_0}{2}\right)^2\left[\int_0^Ldr\,j_l\left(\frac{\omega r}{c_s}\right)\right]^2  \, .
\end{align} 
This can be further approximated as
\begin{align}\label{eqn:no_diff_modulation_2}
	&(\tilde{\xi}^{m}_l(\omega))^2 \approx \left(\frac{\alpha l_p}{2\pi^2 c_s}\right)  \nonumber \\
	\times&\begin{dcases*}
		\frac{\pi}{2}\left(\frac{\omega_0c_s}{2\omega}\right)^2\frac{1}{l}& if $l\ll(\omega L/c_s)$ \\
		\left(C_l\right)^2\left(\frac{\omega L}{c_s}\right)^{2l}(\omega_0L)^2 & if $l\gg(\omega L/c_s)$ \\
	\end{dcases*} \, .
\end{align} 
where we also took the $l\gg 1$ limit with Stirling's approximation in the first entry. Comparing the result \eqref{eqn:total_modulaion_4} from diffraction and the result \eqref{eqn:no_diff_modulation_2} ignoring diffraction,  one immediately observes that for both analyses, the high $l$ modes are severely suppressed, and will not produce any blurring effects. On the other hand, the amplitudes of the low $l$ modes in both analyses scale as $1/l$ with identical scalings in frequencies when $\omega \lesssim \omega_0$, and with numerical factors agreeing up to $\mathcal{O}(1)$.

In Fig.~\ref{fig:diffraction} we plot in red the $|\tilde{\xi}^{m}_l(\omega)|^2$  obtained by numerically integrating Eq.~\eqref{eqn:no_diff_modulation} --- and confirm that this agrees with the 
diffraction results (shown in blue) up to an $\mathcal{O}(1)$ factor. We thus conclude that the diffraction effect does not qualitatively change the level and correlation structure of phase modulation on the telescope's aperture.

\section{Conclusion}
\label{sec:conclusion}

In this paper, we studied the effect of spacetime fluctuations from the VZ effect, specifically modeled by the pixellon field, on the blurring of astrophysical images.  We concluded that it is not constrained by such observations. 
Even though the VZ effect, similar to the previously considered random walk models, leads to a root-mean-squared path length fluctuation of $\sim \sqrt{l_p L}$, we have shown that the transverse correlation between phase modulations on the telescope's aperture will lead to a suppression factor of $D/L$, where $D$ is the size of the aperture.  More specifically, in Sec.~\ref{sec:transverse_correlation}, we used the pixellon model to compute the two-point correlation function of path length fluctuations on the telescope's aperture, and then employed a  Zeta-function regularization technique to obtain UV-independent values for two-point correlations of phase fluctuations of light on the aperture. In Sec.~\ref{sec:blurring_effects}, we applied scalar diffraction theory to convert this level of phase fluctuations to the level of image blurring, confirming that the VZ effect cannot be constrained by image blurring with foreseeable technology. 

In Sec.~\ref{sec:effcts_of_diffraction}, we incorporated the effect of diffraction on light propagation with spacetime fluctuations from the pixellon model.  Unlike in Ref.~\cite{Chen:2014ueo} which considered a spacetime foam model of quantum fluctuations, diffraction does not qualitatively modify the correlation structure of light in the pixellon mode. In this way, diffraction alone does not provide a physical mechanism for the UV cutoff that underlies the Zeta-function regularization carried out in Sec.~\ref{sec:transverse_correlation}.

Note that our results are in contradiction to the claim made by Ref.~\cite{Hogan:2023rea}, which incorrectly analyzed the model of Ref.~\cite{Li:2022mvy} without properly renormalizing the UV divergence.  One concludes that the VZ effect, as motivated by the 't Hooft commutation relations, can give rise to observably large effects in interferometers, while remaining consistent with the observations of images of distant stars.

\acknowledgments

We thank Daine L. Danielson, Temple He, Dongjun Li, Prahar Mitra, Allic Sivaramakrishnan, Jordan Wilson-Gerow, and Yiwen Zhang for helpful discussions. We are supported by the Heising-Simons Foundation ``Observational Signatures of Quantum Gravity'' collaboration grant 2021-2817. In addition, the work of KZ is supported by a Simons Investigator award and the U.S. Department of Energy, Office of Science, Office of High Energy Physics, under Award No. DE-SC0011632. The work of YC is supported by the Brinson Foundation, the Simons Foundation (Award Number 568762) and NSF Grants PHY-2011961 and No. PHY-2011968.

\appendix

\bibliography{bibliography}

\begin{thebibliography}{49}%
\makeatletter
\providecommand \@ifxundefined [1]{%
 \@ifx{#1\undefined}
}%
\providecommand \@ifnum [1]{%
 \ifnum #1\expandafter \@firstoftwo
 \else \expandafter \@secondoftwo
 \fi
}%
\providecommand \@ifx [1]{%
 \ifx #1\expandafter \@firstoftwo
 \else \expandafter \@secondoftwo
 \fi
}%
\providecommand \natexlab [1]{#1}%
\providecommand \enquote  [1]{``#1''}%
\providecommand \bibnamefont  [1]{#1}%
\providecommand \bibfnamefont [1]{#1}%
\providecommand \citenamefont [1]{#1}%
\providecommand \href@noop [0]{\@secondoftwo}%
\providecommand \href [0]{\begingroup \@sanitize@url \@href}%
\providecommand \@href[1]{\@@startlink{#1}\@@href}%
\providecommand \@@href[1]{\endgroup#1\@@endlink}%
\providecommand \@sanitize@url [0]{\catcode `\\12\catcode `\$12\catcode
  `\&12\catcode `\#12\catcode `\^12\catcode `\_12\catcode `\%12\relax}%
\providecommand \@@startlink[1]{}%
\providecommand \@@endlink[0]{}%
\providecommand \url  [0]{\begingroup\@sanitize@url \@url }%
\providecommand \@url [1]{\endgroup\@href {#1}{\urlprefix }}%
\providecommand \urlprefix  [0]{URL }%
\providecommand \Eprint [0]{\href }%
\providecommand \doibase [0]{https://doi.org/}%
\providecommand \selectlanguage [0]{\@gobble}%
\providecommand \bibinfo  [0]{\@secondoftwo}%
\providecommand \bibfield  [0]{\@secondoftwo}%
\providecommand \translation [1]{[#1]}%
\providecommand \BibitemOpen [0]{}%
\providecommand \bibitemStop [0]{}%
\providecommand \bibitemNoStop [0]{.\EOS\space}%
\providecommand \EOS [0]{\spacefactor3000\relax}%
\providecommand \BibitemShut  [1]{\csname bibitem#1\endcsname}%
\let\auto@bib@innerbib\@empty
\bibitem [{\citenamefont {Misner}\ \emph {et~al.}(1973)\citenamefont {Misner},
  \citenamefont {Thorne},\ and\ \citenamefont {Wheeler}}]{Misner_1973}%
  \BibitemOpen
  \bibfield  {author} {\bibinfo {author} {\bibfnamefont {C.~W.}\ \bibnamefont
  {Misner}}, \bibinfo {author} {\bibfnamefont {K.~S.}\ \bibnamefont {Thorne}},\
  and\ \bibinfo {author} {\bibfnamefont {J.~A.}\ \bibnamefont {Wheeler}},\
  }\href@noop {} {\emph {\bibinfo {title} {{Gravitation}}}}\ (\bibinfo
  {publisher} {W. H. Freeman},\ \bibinfo {address} {San Francisco},\ \bibinfo
  {year} {1973})\BibitemShut {NoStop}%
\bibitem [{\citenamefont {Amelino-Camelia}(1994)}]{Amelino-Camelia:1994suz}%
  \BibitemOpen
  \bibfield  {author} {\bibinfo {author} {\bibfnamefont {G.}~\bibnamefont
  {Amelino-Camelia}},\ }\bibfield  {title} {\bibinfo {title} {{Limits on the
  measurability of space-time distances in the semiclassical approximation of
  quantum gravity}},\ }\href {https://doi.org/10.1142/S0217732394003245}
  {\bibfield  {journal} {\bibinfo  {journal} {Mod. Phys. Lett. A}\ }\textbf
  {\bibinfo {volume} {9}},\ \bibinfo {pages} {3415} (\bibinfo {year} {1994})},\
  \Eprint {https://arxiv.org/abs/gr-qc/9603014} {arXiv:gr-qc/9603014}
  \BibitemShut {NoStop}%
\bibitem [{\citenamefont {Diosi}\ and\ \citenamefont
  {Lukacs}(1989)}]{Diosi:1989hy}%
  \BibitemOpen
  \bibfield  {author} {\bibinfo {author} {\bibfnamefont {L.}~\bibnamefont
  {Diosi}}\ and\ \bibinfo {author} {\bibfnamefont {B.}~\bibnamefont {Lukacs}},\
  }\bibfield  {title} {\bibinfo {title} {{On the minimum uncertainty of
  space-time geodesics}},\ }\href
  {https://doi.org/10.1016/0375-9601(89)90375-7} {\bibfield  {journal}
  {\bibinfo  {journal} {Phys. Lett. A}\ }\textbf {\bibinfo {volume} {142}},\
  \bibinfo {pages} {331} (\bibinfo {year} {1989})}\BibitemShut {NoStop}%
\bibitem [{\citenamefont {Wheeler}(1955)}]{Wheeler_1955}%
  \BibitemOpen
  \bibfield  {author} {\bibinfo {author} {\bibfnamefont {J.~A.}\ \bibnamefont
  {Wheeler}},\ }\bibfield  {title} {\bibinfo {title} {Geons},\ }\href
  {https://doi.org/10.1103/PhysRev.97.511} {\bibfield  {journal} {\bibinfo
  {journal} {Phys. Rev.}\ }\textbf {\bibinfo {volume} {97}},\ \bibinfo {pages}
  {511} (\bibinfo {year} {1955})}\BibitemShut {NoStop}%
\bibitem [{\citenamefont {Hawking}\ \emph {et~al.}(1980)\citenamefont
  {Hawking}, \citenamefont {Page},\ and\ \citenamefont
  {Pope}}]{Hawking:1979pi}%
  \BibitemOpen
  \bibfield  {author} {\bibinfo {author} {\bibfnamefont {S.~W.}\ \bibnamefont
  {Hawking}}, \bibinfo {author} {\bibfnamefont {D.~N.}\ \bibnamefont {Page}},\
  and\ \bibinfo {author} {\bibfnamefont {C.~N.}\ \bibnamefont {Pope}},\
  }\bibfield  {title} {\bibinfo {title} {{Quantum Gravitational Bubbles}},\
  }\href {https://doi.org/10.1016/0550-3213(80)90151-0} {\bibfield  {journal}
  {\bibinfo  {journal} {Nucl. Phys. B}\ }\textbf {\bibinfo {volume} {170}},\
  \bibinfo {pages} {283} (\bibinfo {year} {1980})}\BibitemShut {NoStop}%
\bibitem [{\citenamefont {Ashtekar}\ \emph {et~al.}(1992)\citenamefont
  {Ashtekar}, \citenamefont {Rovelli},\ and\ \citenamefont
  {Smolin}}]{Ashtekar_1992}%
  \BibitemOpen
  \bibfield  {author} {\bibinfo {author} {\bibfnamefont {A.}~\bibnamefont
  {Ashtekar}}, \bibinfo {author} {\bibfnamefont {C.}~\bibnamefont {Rovelli}},\
  and\ \bibinfo {author} {\bibfnamefont {L.}~\bibnamefont {Smolin}},\
  }\bibfield  {title} {\bibinfo {title} {Weaving a classical metric with
  quantum threads},\ }\href {https://doi.org/10.1103/physrevlett.69.237}
  {\bibfield  {journal} {\bibinfo  {journal} {Physical Review Letters}\
  }\textbf {\bibinfo {volume} {69}},\ \bibinfo {pages} {237} (\bibinfo {year}
  {1992})}\BibitemShut {NoStop}%
\bibitem [{\citenamefont {Wigner}(1957)}]{Wigner:1957ep}%
  \BibitemOpen
  \bibfield  {author} {\bibinfo {author} {\bibfnamefont {E.~P.}\ \bibnamefont
  {Wigner}},\ }\bibfield  {title} {\bibinfo {title} {{Relativistic Invariance
  and Quantum Phenomena}},\ }\href {https://doi.org/10.1103/RevModPhys.29.255}
  {\bibfield  {journal} {\bibinfo  {journal} {Rev. Mod. Phys.}\ }\textbf
  {\bibinfo {volume} {29}},\ \bibinfo {pages} {255} (\bibinfo {year}
  {1957})}\BibitemShut {NoStop}%
\bibitem [{\citenamefont {Salecker}\ and\ \citenamefont
  {Wigner}(1958)}]{Salecker_1958}%
  \BibitemOpen
  \bibfield  {author} {\bibinfo {author} {\bibfnamefont {H.}~\bibnamefont
  {Salecker}}\ and\ \bibinfo {author} {\bibfnamefont {E.~P.}\ \bibnamefont
  {Wigner}},\ }\bibfield  {title} {\bibinfo {title} {Quantum limitations of the
  measurement of space-time distances},\ }\href
  {https://doi.org/10.1103/PhysRev.109.571} {\bibfield  {journal} {\bibinfo
  {journal} {Phys. Rev.}\ }\textbf {\bibinfo {volume} {109}},\ \bibinfo {pages}
  {571} (\bibinfo {year} {1958})}\BibitemShut {NoStop}%
\bibitem [{\citenamefont {Kwon}\ and\ \citenamefont
  {Hogan}(2016)}]{Kwon:2014yea}%
  \BibitemOpen
  \bibfield  {author} {\bibinfo {author} {\bibfnamefont {O.}~\bibnamefont
  {Kwon}}\ and\ \bibinfo {author} {\bibfnamefont {C.~J.}\ \bibnamefont
  {Hogan}},\ }\bibfield  {title} {\bibinfo {title} {{Interferometric Tests of
  Planckian Quantum Geometry Models}},\ }\href
  {https://doi.org/10.1088/0264-9381/33/10/105004} {\bibfield  {journal}
  {\bibinfo  {journal} {Class. Quant. Grav.}\ }\textbf {\bibinfo {volume}
  {33}},\ \bibinfo {pages} {105004} (\bibinfo {year} {2016})},\ \Eprint
  {https://arxiv.org/abs/1410.8197} {arXiv:1410.8197 [gr-qc]} \BibitemShut
  {NoStop}%
\bibitem [{\citenamefont {Lieu}\ and\ \citenamefont
  {Hillman}(2003)}]{Lieu:2003ee}%
  \BibitemOpen
  \bibfield  {author} {\bibinfo {author} {\bibfnamefont {R.}~\bibnamefont
  {Lieu}}\ and\ \bibinfo {author} {\bibfnamefont {L.~W.}\ \bibnamefont
  {Hillman}},\ }\bibfield  {title} {\bibinfo {title} {{The Phase coherence of
  light from extragalactic sources - direct evidence against first order
  quantum gravity fluctuations in time and space}},\ }\href
  {https://doi.org/10.1086/374350} {\bibfield  {journal} {\bibinfo  {journal}
  {Astrophys. J. Lett.}\ }\textbf {\bibinfo {volume} {585}},\ \bibinfo {pages}
  {L77} (\bibinfo {year} {2003})},\ \Eprint
  {https://arxiv.org/abs/astro-ph/0301184} {arXiv:astro-ph/0301184}
  \BibitemShut {NoStop}%
\bibitem [{\citenamefont {Ng}(2003)}]{Ng:2003jk}%
  \BibitemOpen
  \bibfield  {author} {\bibinfo {author} {\bibfnamefont {Y.~J.}\ \bibnamefont
  {Ng}},\ }\bibfield  {title} {\bibinfo {title} {{Selected topics in Planck
  scale physics}},\ }\href {https://doi.org/10.1142/S0217732303010934}
  {\bibfield  {journal} {\bibinfo  {journal} {Mod. Phys. Lett. A}\ }\textbf
  {\bibinfo {volume} {18}},\ \bibinfo {pages} {1073} (\bibinfo {year}
  {2003})},\ \Eprint {https://arxiv.org/abs/gr-qc/0305019}
  {arXiv:gr-qc/0305019} \BibitemShut {NoStop}%
\bibitem [{\citenamefont {Ragazzoni}\ \emph {et~al.}(2003)\citenamefont
  {Ragazzoni}, \citenamefont {Turatto},\ and\ \citenamefont
  {Gaessler}}]{Ragazzoni:2003tn}%
  \BibitemOpen
  \bibfield  {author} {\bibinfo {author} {\bibfnamefont {R.}~\bibnamefont
  {Ragazzoni}}, \bibinfo {author} {\bibfnamefont {M.}~\bibnamefont {Turatto}},\
  and\ \bibinfo {author} {\bibfnamefont {W.}~\bibnamefont {Gaessler}},\
  }\bibfield  {title} {\bibinfo {title} {{Lack of observational evidence for
  quantum structure of space - time at Planck scales}},\ }\href
  {https://doi.org/10.1086/375046} {\bibfield  {journal} {\bibinfo  {journal}
  {Astrophys. J. Lett.}\ }\textbf {\bibinfo {volume} {587}},\ \bibinfo {pages}
  {L1} (\bibinfo {year} {2003})},\ \Eprint
  {https://arxiv.org/abs/astro-ph/0303043} {arXiv:astro-ph/0303043}
  \BibitemShut {NoStop}%
\bibitem [{\citenamefont {Christiansen}\ \emph {et~al.}(2006)\citenamefont
  {Christiansen}, \citenamefont {Ng},\ and\ \citenamefont {van
  Dam}}]{Christiansen:2005yg}%
  \BibitemOpen
  \bibfield  {author} {\bibinfo {author} {\bibfnamefont {W.~A.}\ \bibnamefont
  {Christiansen}}, \bibinfo {author} {\bibfnamefont {Y.~J.}\ \bibnamefont
  {Ng}},\ and\ \bibinfo {author} {\bibfnamefont {H.}~\bibnamefont {van Dam}},\
  }\bibfield  {title} {\bibinfo {title} {{Probing spacetime foam with
  extragalactic sources}},\ }\href
  {https://doi.org/10.1103/PhysRevLett.96.051301} {\bibfield  {journal}
  {\bibinfo  {journal} {Phys. Rev. Lett.}\ }\textbf {\bibinfo {volume} {96}},\
  \bibinfo {pages} {051301} (\bibinfo {year} {2006})},\ \Eprint
  {https://arxiv.org/abs/gr-qc/0508121} {arXiv:gr-qc/0508121} \BibitemShut
  {NoStop}%
\bibitem [{\citenamefont {Perlman}\ \emph {et~al.}(2011)\citenamefont
  {Perlman}, \citenamefont {Ng}, \citenamefont {Floyd},\ and\ \citenamefont
  {Christiansen}}]{Perlman:2011wv}%
  \BibitemOpen
  \bibfield  {author} {\bibinfo {author} {\bibfnamefont {E.~S.}\ \bibnamefont
  {Perlman}}, \bibinfo {author} {\bibfnamefont {Y.~J.}\ \bibnamefont {Ng}},
  \bibinfo {author} {\bibfnamefont {D.~J.~E.}\ \bibnamefont {Floyd}},\ and\
  \bibinfo {author} {\bibfnamefont {W.~A.}\ \bibnamefont {Christiansen}},\
  }\bibfield  {title} {\bibinfo {title} {{Using Observations of Distant Quasars
  to Constrain Quantum Gravity}},\ }\href
  {https://doi.org/10.1051/0004-6361/201118319} {\bibfield  {journal} {\bibinfo
   {journal} {Astron. Astrophys.}\ }\textbf {\bibinfo {volume} {535}},\
  \bibinfo {pages} {L9} (\bibinfo {year} {2011})},\ \Eprint
  {https://arxiv.org/abs/1110.4986} {arXiv:1110.4986 [astro-ph.CO]}
  \BibitemShut {NoStop}%
\bibitem [{\citenamefont {Perlman}\ \emph {et~al.}(2015)\citenamefont
  {Perlman}, \citenamefont {Rappaport}, \citenamefont {Christiansen},
  \citenamefont {Ng}, \citenamefont {DeVore},\ and\ \citenamefont
  {Pooley}}]{Perlman:2014cwa}%
  \BibitemOpen
  \bibfield  {author} {\bibinfo {author} {\bibfnamefont {E.~S.}\ \bibnamefont
  {Perlman}}, \bibinfo {author} {\bibfnamefont {S.~A.}\ \bibnamefont
  {Rappaport}}, \bibinfo {author} {\bibfnamefont {W.~A.}\ \bibnamefont
  {Christiansen}}, \bibinfo {author} {\bibfnamefont {Y.~J.}\ \bibnamefont
  {Ng}}, \bibinfo {author} {\bibfnamefont {J.}~\bibnamefont {DeVore}},\ and\
  \bibinfo {author} {\bibfnamefont {D.}~\bibnamefont {Pooley}},\ }\bibfield
  {title} {\bibinfo {title} {{New Constraints on Quantum Gravity from X-ray and
  Gamma-Ray Observations}},\ }\href
  {https://doi.org/10.1088/0004-637X/805/1/10} {\bibfield  {journal} {\bibinfo
  {journal} {Astrophys. J.}\ }\textbf {\bibinfo {volume} {805}},\ \bibinfo
  {pages} {10} (\bibinfo {year} {2015})},\ \Eprint
  {https://arxiv.org/abs/1411.7262} {arXiv:1411.7262 [astro-ph.CO]}
  \BibitemShut {NoStop}%
\bibitem [{\citenamefont {Hogan}\ \emph {et~al.}(2023)\citenamefont {Hogan},
  \citenamefont {Kwon},\ and\ \citenamefont {Selub}}]{Hogan:2023rea}%
  \BibitemOpen
  \bibfield  {author} {\bibinfo {author} {\bibfnamefont {C.}~\bibnamefont
  {Hogan}}, \bibinfo {author} {\bibfnamefont {O.}~\bibnamefont {Kwon}},\ and\
  \bibinfo {author} {\bibfnamefont {N.}~\bibnamefont {Selub}},\ }\bibfield
  {title} {\bibinfo {title} {{Angular spectrum of quantum fluctuations in
  causal structure}},\ }\href@noop {} {\  (\bibinfo {year} {2023})},\ \Eprint
  {https://arxiv.org/abs/2303.06563} {arXiv:2303.06563 [gr-qc]} \BibitemShut
  {NoStop}%
\bibitem [{\citenamefont {Steinbring}(2023)}]{Steinbring:2023abf}%
  \BibitemOpen
  \bibfield  {author} {\bibinfo {author} {\bibfnamefont {E.}~\bibnamefont
  {Steinbring}},\ }\bibfield  {title} {\bibinfo {title} {{Holographic
  Quantum-Foam Blurring Is Consistent with Observations of Gamma-Ray Burst
  GRB221009A}},\ }\href {https://doi.org/10.3390/galaxies11060115} {\bibfield
  {journal} {\bibinfo  {journal} {Galaxies}\ }\textbf {\bibinfo {volume}
  {11}},\ \bibinfo {pages} {115} (\bibinfo {year} {2023})}\BibitemShut
  {NoStop}%
\bibitem [{\citenamefont {Thorne}\ and\ \citenamefont
  {Blandford}(2017)}]{thorne2017modern}%
  \BibitemOpen
  \bibfield  {author} {\bibinfo {author} {\bibfnamefont {K.~S.}\ \bibnamefont
  {Thorne}}\ and\ \bibinfo {author} {\bibfnamefont {R.~D.}\ \bibnamefont
  {Blandford}},\ }\href@noop {} {\emph {\bibinfo {title} {Modern classical
  physics: optics, fluids, plasmas, elasticity, relativity, and statistical
  physics}}}\ (\bibinfo  {publisher} {Princeton University Press},\ \bibinfo
  {year} {2017})\BibitemShut {NoStop}%
\bibitem [{\citenamefont {Chen}\ \emph {et~al.}(2014)\citenamefont {Chen},
  \citenamefont {Wen},\ and\ \citenamefont {Ma}}]{Chen:2014ueo}%
  \BibitemOpen
  \bibfield  {author} {\bibinfo {author} {\bibfnamefont {Y.}~\bibnamefont
  {Chen}}, \bibinfo {author} {\bibfnamefont {L.}~\bibnamefont {Wen}},\ and\
  \bibinfo {author} {\bibfnamefont {Y.}~\bibnamefont {Ma}},\ }\bibfield
  {title} {\bibinfo {title} {{Photons with sub-Planckian energy cannot
  efficiently probe space-time foam}},\ }\href
  {https://doi.org/10.1103/PhysRevD.90.063011} {\bibfield  {journal} {\bibinfo
  {journal} {Phys. Rev. D}\ }\textbf {\bibinfo {volume} {90}},\ \bibinfo
  {pages} {063011} (\bibinfo {year} {2014})},\ \Eprint
  {https://arxiv.org/abs/1504.07509} {arXiv:1504.07509 [astro-ph.HE]}
  \BibitemShut {NoStop}%
\bibitem [{\citenamefont {Verlinde}\ and\ \citenamefont {Zurek}(2021)}]{VZ1}%
  \BibitemOpen
  \bibfield  {author} {\bibinfo {author} {\bibfnamefont {E.~P.}\ \bibnamefont
  {Verlinde}}\ and\ \bibinfo {author} {\bibfnamefont {K.~M.}\ \bibnamefont
  {Zurek}},\ }\bibfield  {title} {\bibinfo {title} {{Observational signatures
  of quantum gravity in interferometers}},\ }\href
  {https://doi.org/10.1016/j.physletb.2021.136663} {\bibfield  {journal}
  {\bibinfo  {journal} {Phys. Lett. B}\ }\textbf {\bibinfo {volume} {822}},\
  \bibinfo {pages} {136663} (\bibinfo {year} {2021})},\ \Eprint
  {https://arxiv.org/abs/1902.08207} {arXiv:1902.08207 [gr-qc]} \BibitemShut
  {NoStop}%
\bibitem [{\citenamefont {Verlinde}\ and\ \citenamefont {Zurek}(2020)}]{VZ2}%
  \BibitemOpen
  \bibfield  {author} {\bibinfo {author} {\bibfnamefont {E.}~\bibnamefont
  {Verlinde}}\ and\ \bibinfo {author} {\bibfnamefont {K.~M.}\ \bibnamefont
  {Zurek}},\ }\bibfield  {title} {\bibinfo {title} {{Spacetime Fluctuations in
  AdS/CFT}},\ }\href {https://doi.org/10.1007/JHEP04(2020)209} {\bibfield
  {journal} {\bibinfo  {journal} {JHEP}\ }\textbf {\bibinfo {volume} {04}},\
  \bibinfo {pages} {209}},\ \Eprint {https://arxiv.org/abs/1911.02018}
  {arXiv:1911.02018 [hep-th]} \BibitemShut {NoStop}%
\bibitem [{\citenamefont {Banks}\ and\ \citenamefont
  {Zurek}(2021)}]{Banks:2021jwj}%
  \BibitemOpen
  \bibfield  {author} {\bibinfo {author} {\bibfnamefont {T.}~\bibnamefont
  {Banks}}\ and\ \bibinfo {author} {\bibfnamefont {K.~M.}\ \bibnamefont
  {Zurek}},\ }\bibfield  {title} {\bibinfo {title} {{Conformal description of
  near-horizon vacuum states}},\ }\href
  {https://doi.org/10.1103/PhysRevD.104.126026} {\bibfield  {journal} {\bibinfo
   {journal} {Phys. Rev. D}\ }\textbf {\bibinfo {volume} {104}},\ \bibinfo
  {pages} {126026} (\bibinfo {year} {2021})},\ \Eprint
  {https://arxiv.org/abs/2108.04806} {arXiv:2108.04806 [hep-th]} \BibitemShut
  {NoStop}%
\bibitem [{\citenamefont {Gukov}\ \emph {et~al.}(2023)\citenamefont {Gukov},
  \citenamefont {Lee},\ and\ \citenamefont {Zurek}}]{Gukov:2022oed}%
  \BibitemOpen
  \bibfield  {author} {\bibinfo {author} {\bibfnamefont {S.}~\bibnamefont
  {Gukov}}, \bibinfo {author} {\bibfnamefont {V.~S.~H.}\ \bibnamefont {Lee}},\
  and\ \bibinfo {author} {\bibfnamefont {K.~M.}\ \bibnamefont {Zurek}},\
  }\bibfield  {title} {\bibinfo {title} {{Near-horizon quantum dynamics of 4D
  Einstein gravity from 2D Jackiw-Teitelboim gravity}},\ }\href
  {https://doi.org/10.1103/PhysRevD.107.016004} {\bibfield  {journal} {\bibinfo
   {journal} {Phys. Rev. D}\ }\textbf {\bibinfo {volume} {107}},\ \bibinfo
  {pages} {016004} (\bibinfo {year} {2023})},\ \Eprint
  {https://arxiv.org/abs/2205.02233} {arXiv:2205.02233 [hep-th]} \BibitemShut
  {NoStop}%
\bibitem [{\citenamefont {Verlinde}\ and\ \citenamefont {Zurek}(2022)}]{VZ3}%
  \BibitemOpen
  \bibfield  {author} {\bibinfo {author} {\bibfnamefont {E.}~\bibnamefont
  {Verlinde}}\ and\ \bibinfo {author} {\bibfnamefont {K.~M.}\ \bibnamefont
  {Zurek}},\ }\bibfield  {title} {\bibinfo {title} {{Modular fluctuations from
  shockwave geometries}},\ }\href {https://doi.org/10.1103/PhysRevD.106.106011}
  {\bibfield  {journal} {\bibinfo  {journal} {Phys. Rev. D}\ }\textbf {\bibinfo
  {volume} {106}},\ \bibinfo {pages} {106011} (\bibinfo {year} {2022})},\
  \Eprint {https://arxiv.org/abs/2208.01059} {arXiv:2208.01059 [hep-th]}
  \BibitemShut {NoStop}%
\bibitem [{\citenamefont {'t~Hooft}(1996)}]{tHooft:1996rdg}%
  \BibitemOpen
  \bibfield  {author} {\bibinfo {author} {\bibfnamefont {G.}~\bibnamefont
  {'t~Hooft}},\ }\bibfield  {title} {\bibinfo {title} {{The Scattering matrix
  approach for the quantum black hole: An Overview}},\ }\href
  {https://doi.org/10.1142/S0217751X96002145} {\bibfield  {journal} {\bibinfo
  {journal} {Int. J. Mod. Phys. A}\ }\textbf {\bibinfo {volume} {11}},\
  \bibinfo {pages} {4623} (\bibinfo {year} {1996})},\ \Eprint
  {https://arxiv.org/abs/gr-qc/9607022} {arXiv:gr-qc/9607022} \BibitemShut
  {NoStop}%
\bibitem [{\citenamefont {He}\ \emph {et~al.}(2023)\citenamefont {He},
  \citenamefont {Raclariu},\ and\ \citenamefont {Zurek}}]{He:2023qha}%
  \BibitemOpen
  \bibfield  {author} {\bibinfo {author} {\bibfnamefont {T.}~\bibnamefont
  {He}}, \bibinfo {author} {\bibfnamefont {A.-M.}\ \bibnamefont {Raclariu}},\
  and\ \bibinfo {author} {\bibfnamefont {K.~M.}\ \bibnamefont {Zurek}},\
  }\bibfield  {title} {\bibinfo {title} {{From Shockwaves to the Gravitational
  Memory Effect}},\ }\href@noop {} {\  (\bibinfo {year} {2023})},\ \Eprint
  {https://arxiv.org/abs/2305.14411} {arXiv:2305.14411 [hep-th]} \BibitemShut
  {NoStop}%
\bibitem [{\citenamefont {Dray}\ and\ \citenamefont
  {'t~Hooft}(1985{\natexlab{a}})}]{Dray:1984ha}%
  \BibitemOpen
  \bibfield  {author} {\bibinfo {author} {\bibfnamefont {T.}~\bibnamefont
  {Dray}}\ and\ \bibinfo {author} {\bibfnamefont {G.}~\bibnamefont
  {'t~Hooft}},\ }\bibfield  {title} {\bibinfo {title} {{The Gravitational Shock
  Wave of a Massless Particle}},\ }\href
  {https://doi.org/10.1016/0550-3213(85)90525-5} {\bibfield  {journal}
  {\bibinfo  {journal} {Nucl. Phys. B}\ }\textbf {\bibinfo {volume} {253}},\
  \bibinfo {pages} {173} (\bibinfo {year} {1985}{\natexlab{a}})}\BibitemShut
  {NoStop}%
\bibitem [{\citenamefont {Dray}\ and\ \citenamefont
  {'t~Hooft}(1985{\natexlab{b}})}]{Dray:1985yt}%
  \BibitemOpen
  \bibfield  {author} {\bibinfo {author} {\bibfnamefont {T.}~\bibnamefont
  {Dray}}\ and\ \bibinfo {author} {\bibfnamefont {G.}~\bibnamefont
  {'t~Hooft}},\ }\bibfield  {title} {\bibinfo {title} {{The Effect of Spherical
  Shells of Matter on the Schwarzschild Black Hole}},\ }\href
  {https://doi.org/10.1007/BF01215912} {\bibfield  {journal} {\bibinfo
  {journal} {Commun. Math. Phys.}\ }\textbf {\bibinfo {volume} {99}},\ \bibinfo
  {pages} {613} (\bibinfo {year} {1985}{\natexlab{b}})}\BibitemShut {NoStop}%
\bibitem [{\citenamefont {'t~Hooft}(2018)}]{tHooft:2018fxg}%
  \BibitemOpen
  \bibfield  {author} {\bibinfo {author} {\bibfnamefont {G.}~\bibnamefont
  {'t~Hooft}},\ }\bibfield  {title} {\bibinfo {title} {{Discreteness of Black
  Hole Microstates}},\ }\href@noop {} {\  (\bibinfo {year} {2018})},\ \Eprint
  {https://arxiv.org/abs/1809.05367} {arXiv:1809.05367 [gr-qc]} \BibitemShut
  {NoStop}%
\bibitem [{\citenamefont {Mertens}\ \emph {et~al.}(2017)\citenamefont
  {Mertens}, \citenamefont {Turiaci},\ and\ \citenamefont
  {Verlinde}}]{Mertens:2017mtv}%
  \BibitemOpen
  \bibfield  {author} {\bibinfo {author} {\bibfnamefont {T.~G.}\ \bibnamefont
  {Mertens}}, \bibinfo {author} {\bibfnamefont {G.~J.}\ \bibnamefont
  {Turiaci}},\ and\ \bibinfo {author} {\bibfnamefont {H.~L.}\ \bibnamefont
  {Verlinde}},\ }\bibfield  {title} {\bibinfo {title} {{Solving the Schwarzian
  via the Conformal Bootstrap}},\ }\href
  {https://doi.org/10.1007/JHEP08(2017)136} {\bibfield  {journal} {\bibinfo
  {journal} {JHEP}\ }\textbf {\bibinfo {volume} {08}},\ \bibinfo {pages}
  {136}},\ \Eprint {https://arxiv.org/abs/1705.08408} {arXiv:1705.08408
  [hep-th]} \BibitemShut {NoStop}%
\bibitem [{\citenamefont {Lam}\ \emph {et~al.}(2018)\citenamefont {Lam},
  \citenamefont {Mertens}, \citenamefont {Turiaci},\ and\ \citenamefont
  {Verlinde}}]{Lam:2018pvp}%
  \BibitemOpen
  \bibfield  {author} {\bibinfo {author} {\bibfnamefont {H.~T.}\ \bibnamefont
  {Lam}}, \bibinfo {author} {\bibfnamefont {T.~G.}\ \bibnamefont {Mertens}},
  \bibinfo {author} {\bibfnamefont {G.~J.}\ \bibnamefont {Turiaci}},\ and\
  \bibinfo {author} {\bibfnamefont {H.}~\bibnamefont {Verlinde}},\ }\bibfield
  {title} {\bibinfo {title} {{Shockwave S-matrix from Schwarzian Quantum
  Mechanics}},\ }\href {https://doi.org/10.1007/JHEP11(2018)182} {\bibfield
  {journal} {\bibinfo  {journal} {JHEP}\ }\textbf {\bibinfo {volume} {11}},\
  \bibinfo {pages} {182}},\ \Eprint {https://arxiv.org/abs/1804.09834}
  {arXiv:1804.09834 [hep-th]} \BibitemShut {NoStop}%
\bibitem [{\citenamefont {Ng}\ and\ \citenamefont {van Dam}(2000)}]{Ng:1999hm}%
  \BibitemOpen
  \bibfield  {author} {\bibinfo {author} {\bibfnamefont {Y.~J.}\ \bibnamefont
  {Ng}}\ and\ \bibinfo {author} {\bibfnamefont {H.}~\bibnamefont {van Dam}},\
  }\bibfield  {title} {\bibinfo {title} {{Measuring the foaminess of space-time
  with gravity - wave interferometers}},\ }\href
  {https://doi.org/10.1023/A:1003745212871} {\bibfield  {journal} {\bibinfo
  {journal} {Found. Phys.}\ }\textbf {\bibinfo {volume} {30}},\ \bibinfo
  {pages} {795} (\bibinfo {year} {2000})},\ \Eprint
  {https://arxiv.org/abs/gr-qc/9906003} {arXiv:gr-qc/9906003} \BibitemShut
  {NoStop}%
\bibitem [{\citenamefont {Hogan}(2008)}]{Hogan:2007pk}%
  \BibitemOpen
  \bibfield  {author} {\bibinfo {author} {\bibfnamefont {C.~J.}\ \bibnamefont
  {Hogan}},\ }\bibfield  {title} {\bibinfo {title} {{Measurement of Quantum
  Fluctuations in Geometry}},\ }\href
  {https://doi.org/10.1103/PhysRevD.77.104031} {\bibfield  {journal} {\bibinfo
  {journal} {Phys. Rev. D}\ }\textbf {\bibinfo {volume} {77}},\ \bibinfo
  {pages} {104031} (\bibinfo {year} {2008})},\ \Eprint
  {https://arxiv.org/abs/0712.3419} {arXiv:0712.3419 [gr-qc]} \BibitemShut
  {NoStop}%
\bibitem [{\citenamefont {Zurek}(2022{\natexlab{a}})}]{Zurek:2020ukz}%
  \BibitemOpen
  \bibfield  {author} {\bibinfo {author} {\bibfnamefont {K.~M.}\ \bibnamefont
  {Zurek}},\ }\bibfield  {title} {\bibinfo {title} {{On vacuum fluctuations in
  quantum gravity and interferometer arm fluctuations}},\ }\href
  {https://doi.org/10.1016/j.physletb.2022.136910} {\bibfield  {journal}
  {\bibinfo  {journal} {Phys. Lett. B}\ }\textbf {\bibinfo {volume} {826}},\
  \bibinfo {pages} {136910} (\bibinfo {year} {2022}{\natexlab{a}})},\ \Eprint
  {https://arxiv.org/abs/2012.05870} {arXiv:2012.05870 [hep-th]} \BibitemShut
  {NoStop}%
\bibitem [{\citenamefont {Li}\ \emph {et~al.}(2023)\citenamefont {Li},
  \citenamefont {Lee}, \citenamefont {Chen},\ and\ \citenamefont
  {Zurek}}]{Li:2022mvy}%
  \BibitemOpen
  \bibfield  {author} {\bibinfo {author} {\bibfnamefont {D.}~\bibnamefont
  {Li}}, \bibinfo {author} {\bibfnamefont {V.~S.~H.}\ \bibnamefont {Lee}},
  \bibinfo {author} {\bibfnamefont {Y.}~\bibnamefont {Chen}},\ and\ \bibinfo
  {author} {\bibfnamefont {K.~M.}\ \bibnamefont {Zurek}},\ }\bibfield  {title}
  {\bibinfo {title} {{Interferometer response to geontropic fluctuations}},\
  }\href {https://doi.org/10.1103/PhysRevD.107.024002} {\bibfield  {journal}
  {\bibinfo  {journal} {Phys. Rev. D}\ }\textbf {\bibinfo {volume} {107}},\
  \bibinfo {pages} {024002} (\bibinfo {year} {2023})},\ \Eprint
  {https://arxiv.org/abs/2209.07543} {arXiv:2209.07543 [gr-qc]} \BibitemShut
  {NoStop}%
\bibitem [{\citenamefont {Zurek}(2022{\natexlab{b}})}]{Zurek:2022xzl}%
  \BibitemOpen
  \bibfield  {author} {\bibinfo {author} {\bibfnamefont {K.~M.}\ \bibnamefont
  {Zurek}},\ }\bibfield  {title} {\bibinfo {title} {{Snowmass 2021 White Paper:
  Observational Signatures of Quantum Gravity}},\ }\href@noop {} {\  (\bibinfo
  {year} {2022}{\natexlab{b}})},\ \Eprint {https://arxiv.org/abs/2205.01799}
  {arXiv:2205.01799 [gr-qc]} \BibitemShut {NoStop}%
\bibitem [{\citenamefont {Perlman}\ \emph {et~al.}(2017)\citenamefont
  {Perlman}, \citenamefont {Rappaport}, \citenamefont {Ng}, \citenamefont
  {Christiansen}, \citenamefont {DeVore},\ and\ \citenamefont
  {Pooley}}]{Perlman:2016xbc}%
  \BibitemOpen
  \bibfield  {author} {\bibinfo {author} {\bibfnamefont {E.~S.}\ \bibnamefont
  {Perlman}}, \bibinfo {author} {\bibfnamefont {S.~A.}\ \bibnamefont
  {Rappaport}}, \bibinfo {author} {\bibfnamefont {Y.~J.}\ \bibnamefont {Ng}},
  \bibinfo {author} {\bibfnamefont {W.~A.}\ \bibnamefont {Christiansen}},
  \bibinfo {author} {\bibfnamefont {J.}~\bibnamefont {DeVore}},\ and\ \bibinfo
  {author} {\bibfnamefont {D.}~\bibnamefont {Pooley}},\ }\bibfield  {title}
  {\bibinfo {title} {{New constraints on quantum foam models from X-ray and
  gamma-ray observations of distant quasars}},\ }in\ \href
  {https://doi.org/10.1142/9789813226609_0523} {\emph {\bibinfo {booktitle}
  {{14th Marcel Grossmann Meeting on Recent Developments in Theoretical and
  Experimental General Relativity, Astrophysics, and Relativistic Field
  Theories}}}},\ Vol.~\bibinfo {volume} {4}\ (\bibinfo {year} {2017})\ pp.\
  \bibinfo {pages} {3935--3941},\ \Eprint {https://arxiv.org/abs/1607.08551}
  {arXiv:1607.08551 [astro-ph.CO]} \BibitemShut {NoStop}%
\bibitem [{\citenamefont {Christiansen}\ \emph {et~al.}(2011)\citenamefont
  {Christiansen}, \citenamefont {Ng}, \citenamefont {Floyd},\ and\
  \citenamefont {Perlman}}]{Christiansen:2009bz}%
  \BibitemOpen
  \bibfield  {author} {\bibinfo {author} {\bibfnamefont {W.~A.}\ \bibnamefont
  {Christiansen}}, \bibinfo {author} {\bibfnamefont {Y.~J.}\ \bibnamefont
  {Ng}}, \bibinfo {author} {\bibfnamefont {D.~J.~E.}\ \bibnamefont {Floyd}},\
  and\ \bibinfo {author} {\bibfnamefont {E.~S.}\ \bibnamefont {Perlman}},\
  }\bibfield  {title} {\bibinfo {title} {{Limits on Spacetime Foam}},\ }\href
  {https://doi.org/10.1103/PhysRevD.83.084003} {\bibfield  {journal} {\bibinfo
  {journal} {Phys. Rev. D}\ }\textbf {\bibinfo {volume} {83}},\ \bibinfo
  {pages} {084003} (\bibinfo {year} {2011})},\ \Eprint
  {https://arxiv.org/abs/0912.0535} {arXiv:0912.0535 [astro-ph.CO]}
  \BibitemShut {NoStop}%
\bibitem [{\citenamefont {Beckwith}\ \emph {et~al.}(2006)\citenamefont
  {Beckwith} \emph {et~al.}}]{Beckwith:2006qi}%
  \BibitemOpen
  \bibfield  {author} {\bibinfo {author} {\bibfnamefont {S.~V.~W.}\
  \bibnamefont {Beckwith}} \emph {et~al.},\ }\bibfield  {title} {\bibinfo
  {title} {{The Hubble Ultra Deep Field}},\ }\href
  {https://doi.org/10.1086/507302} {\bibfield  {journal} {\bibinfo  {journal}
  {Astron. J.}\ }\textbf {\bibinfo {volume} {132}},\ \bibinfo {pages} {1729}
  (\bibinfo {year} {2006})},\ \Eprint {https://arxiv.org/abs/astro-ph/0607632}
  {arXiv:astro-ph/0607632} \BibitemShut {NoStop}%
\bibitem [{\citenamefont {{Welch}}\ \emph {et~al.}(2022)\citenamefont
  {{Welch}}, \citenamefont {{Coe}}, \citenamefont {{Diego}}, \citenamefont
  {{Zitrin}}, \citenamefont {{Zackrisson}}, \citenamefont {{Dimauro}},
  \citenamefont {{Jim{\'e}nez-Teja}}, \citenamefont {{Kelly}}, \citenamefont
  {{Mahler}}, \citenamefont {{Oguri}}, \citenamefont {{Timmes}}, \citenamefont
  {{Windhorst}}, \citenamefont {{Florian}}, \citenamefont {{de Mink}},
  \citenamefont {{Avila}}, \citenamefont {{Anderson}}, \citenamefont
  {{Bradley}}, \citenamefont {{Sharon}}, \citenamefont {{Vikaeus}},
  \citenamefont {{McCandliss}}, \citenamefont {{Brada{\v{c}}}}, \citenamefont
  {{Rigby}}, \citenamefont {{Frye}}, \citenamefont {{Toft}}, \citenamefont
  {{Strait}}, \citenamefont {{Trenti}}, \citenamefont {{Sharma}}, \citenamefont
  {{Andrade-Santos}},\ and\ \citenamefont {{Broadhurst}}}]{Welch_2022}%
  \BibitemOpen
  \bibfield  {author} {\bibinfo {author} {\bibfnamefont {B.}~\bibnamefont
  {{Welch}}}, \bibinfo {author} {\bibfnamefont {D.}~\bibnamefont {{Coe}}},
  \bibinfo {author} {\bibfnamefont {J.~M.}\ \bibnamefont {{Diego}}}, \bibinfo
  {author} {\bibfnamefont {A.}~\bibnamefont {{Zitrin}}}, \bibinfo {author}
  {\bibfnamefont {E.}~\bibnamefont {{Zackrisson}}}, \bibinfo {author}
  {\bibfnamefont {P.}~\bibnamefont {{Dimauro}}}, \bibinfo {author}
  {\bibfnamefont {Y.}~\bibnamefont {{Jim{\'e}nez-Teja}}}, \bibinfo {author}
  {\bibfnamefont {P.}~\bibnamefont {{Kelly}}}, \bibinfo {author} {\bibfnamefont
  {G.}~\bibnamefont {{Mahler}}}, \bibinfo {author} {\bibfnamefont
  {M.}~\bibnamefont {{Oguri}}}, \bibinfo {author} {\bibfnamefont {F.~X.}\
  \bibnamefont {{Timmes}}}, \bibinfo {author} {\bibfnamefont {R.}~\bibnamefont
  {{Windhorst}}}, \bibinfo {author} {\bibfnamefont {M.}~\bibnamefont
  {{Florian}}}, \bibinfo {author} {\bibfnamefont {S.~E.}\ \bibnamefont {{de
  Mink}}}, \bibinfo {author} {\bibfnamefont {R.~J.}\ \bibnamefont {{Avila}}},
  \bibinfo {author} {\bibfnamefont {J.}~\bibnamefont {{Anderson}}}, \bibinfo
  {author} {\bibfnamefont {L.}~\bibnamefont {{Bradley}}}, \bibinfo {author}
  {\bibfnamefont {K.}~\bibnamefont {{Sharon}}}, \bibinfo {author}
  {\bibfnamefont {A.}~\bibnamefont {{Vikaeus}}}, \bibinfo {author}
  {\bibfnamefont {S.}~\bibnamefont {{McCandliss}}}, \bibinfo {author}
  {\bibfnamefont {M.}~\bibnamefont {{Brada{\v{c}}}}}, \bibinfo {author}
  {\bibfnamefont {J.}~\bibnamefont {{Rigby}}}, \bibinfo {author} {\bibfnamefont
  {B.}~\bibnamefont {{Frye}}}, \bibinfo {author} {\bibfnamefont
  {S.}~\bibnamefont {{Toft}}}, \bibinfo {author} {\bibfnamefont
  {V.}~\bibnamefont {{Strait}}}, \bibinfo {author} {\bibfnamefont
  {M.}~\bibnamefont {{Trenti}}}, \bibinfo {author} {\bibfnamefont
  {S.}~\bibnamefont {{Sharma}}}, \bibinfo {author} {\bibfnamefont
  {F.}~\bibnamefont {{Andrade-Santos}}},\ and\ \bibinfo {author} {\bibfnamefont
  {T.}~\bibnamefont {{Broadhurst}}},\ }\bibfield  {title} {\bibinfo {title} {{A
  highly magnified star at redshift 6.2}},\ }\href
  {https://doi.org/10.1038/s41586-022-04449-y} {\bibfield  {journal} {\bibinfo
  {journal} {Nature}\ }\textbf {\bibinfo {volume} {603}},\ \bibinfo {pages}
  {815} (\bibinfo {year} {2022})}\BibitemShut {NoStop}%
\bibitem [{\citenamefont {Coule}(2005)}]{Coule:2004qf}%
  \BibitemOpen
  \bibfield  {author} {\bibinfo {author} {\bibfnamefont {D.~H.}\ \bibnamefont
  {Coule}},\ }\bibfield  {title} {\bibinfo {title} {{Quantum cosmological
  models}},\ }\href {https://doi.org/10.1088/0264-9381/22/12/R02} {\bibfield
  {journal} {\bibinfo  {journal} {Class. Quant. Grav.}\ }\textbf {\bibinfo
  {volume} {22}},\ \bibinfo {pages} {R125} (\bibinfo {year} {2005})},\ \Eprint
  {https://arxiv.org/abs/gr-qc/0412026} {arXiv:gr-qc/0412026} \BibitemShut
  {NoStop}%
\bibitem [{\citenamefont {Dowker}\ \emph {et~al.}(2010)\citenamefont {Dowker},
  \citenamefont {Henson},\ and\ \citenamefont {Sorkin}}]{Dowker:2010pf}%
  \BibitemOpen
  \bibfield  {author} {\bibinfo {author} {\bibfnamefont {F.}~\bibnamefont
  {Dowker}}, \bibinfo {author} {\bibfnamefont {J.}~\bibnamefont {Henson}},\
  and\ \bibinfo {author} {\bibfnamefont {R.}~\bibnamefont {Sorkin}},\
  }\bibfield  {title} {\bibinfo {title} {{Discreteness and the transmission of
  light from distant sources}},\ }\href
  {https://doi.org/10.1103/PhysRevD.82.104048} {\bibfield  {journal} {\bibinfo
  {journal} {Phys. Rev. D}\ }\textbf {\bibinfo {volume} {82}},\ \bibinfo
  {pages} {104048} (\bibinfo {year} {2010})},\ \Eprint
  {https://arxiv.org/abs/1009.3058} {arXiv:1009.3058 [gr-qc]} \BibitemShut
  {NoStop}%
\bibitem [{\citenamefont {Jackson}(1998)}]{Jackson:1998nia}%
  \BibitemOpen
  \bibfield  {author} {\bibinfo {author} {\bibfnamefont {J.~D.}\ \bibnamefont
  {Jackson}},\ }\href@noop {} {\emph {\bibinfo {title} {{Classical
  Electrodynamics}}}}\ (\bibinfo  {publisher} {Wiley},\ \bibinfo {year}
  {1998})\BibitemShut {NoStop}%
\bibitem [{\citenamefont {Zhang}\ and\ \citenamefont
  {Zurek}(2023)}]{Zhang:2023mkf}%
  \BibitemOpen
  \bibfield  {author} {\bibinfo {author} {\bibfnamefont {Y.}~\bibnamefont
  {Zhang}}\ and\ \bibinfo {author} {\bibfnamefont {K.~M.}\ \bibnamefont
  {Zurek}},\ }\bibfield  {title} {\bibinfo {title} {{Stochastic description of
  near-horizon fluctuations in Rindler-AdS}},\ }\href
  {https://doi.org/10.1103/PhysRevD.108.066002} {\bibfield  {journal} {\bibinfo
   {journal} {Phys. Rev. D}\ }\textbf {\bibinfo {volume} {108}},\ \bibinfo
  {pages} {066002} (\bibinfo {year} {2023})},\ \Eprint
  {https://arxiv.org/abs/2304.12349} {arXiv:2304.12349 [hep-th]} \BibitemShut
  {NoStop}%
\bibitem [{\citenamefont {Bub}\ \emph {et~al.}(2023)\citenamefont {Bub},
  \citenamefont {Chen}, \citenamefont {Du}, \citenamefont {Li}, \citenamefont
  {Zhang},\ and\ \citenamefont {Zurek}}]{Bub:2023bfi}%
  \BibitemOpen
  \bibfield  {author} {\bibinfo {author} {\bibfnamefont {M.~W.}\ \bibnamefont
  {Bub}}, \bibinfo {author} {\bibfnamefont {Y.}~\bibnamefont {Chen}}, \bibinfo
  {author} {\bibfnamefont {Y.}~\bibnamefont {Du}}, \bibinfo {author}
  {\bibfnamefont {D.}~\bibnamefont {Li}}, \bibinfo {author} {\bibfnamefont
  {Y.}~\bibnamefont {Zhang}},\ and\ \bibinfo {author} {\bibfnamefont {K.~M.}\
  \bibnamefont {Zurek}},\ }\bibfield  {title} {\bibinfo {title} {{Quantum
  gravity background in next-generation gravitational wave detectors}},\ }\href
  {https://doi.org/10.1103/PhysRevD.108.064038} {\bibfield  {journal} {\bibinfo
   {journal} {Phys. Rev. D}\ }\textbf {\bibinfo {volume} {108}},\ \bibinfo
  {pages} {064038} (\bibinfo {year} {2023})},\ \Eprint
  {https://arxiv.org/abs/2305.11224} {arXiv:2305.11224 [gr-qc]} \BibitemShut
  {NoStop}%
\bibitem [{\citenamefont {Casimir}\ and\ \citenamefont
  {Polder}(1948)}]{Casimir_1948}%
  \BibitemOpen
  \bibfield  {author} {\bibinfo {author} {\bibfnamefont {H.~B.~G.}\
  \bibnamefont {Casimir}}\ and\ \bibinfo {author} {\bibfnamefont
  {D.}~\bibnamefont {Polder}},\ }\bibfield  {title} {\bibinfo {title} {The
  influence of retardation on the london-van der waals forces},\ }\href
  {https://doi.org/10.1103/PhysRev.73.360} {\bibfield  {journal} {\bibinfo
  {journal} {Phys. Rev.}\ }\textbf {\bibinfo {volume} {73}},\ \bibinfo {pages}
  {360} (\bibinfo {year} {1948})}\BibitemShut {NoStop}%
\bibitem [{\citenamefont {Hawking}(1977)}]{Hawking:1976ja}%
  \BibitemOpen
  \bibfield  {author} {\bibinfo {author} {\bibfnamefont {S.~W.}\ \bibnamefont
  {Hawking}},\ }\bibfield  {title} {\bibinfo {title} {{Zeta Function
  Regularization of Path Integrals in Curved Space-Time}},\ }\href
  {https://doi.org/10.1007/BF01626516} {\bibfield  {journal} {\bibinfo
  {journal} {Commun. Math. Phys.}\ }\textbf {\bibinfo {volume} {55}},\ \bibinfo
  {pages} {133} (\bibinfo {year} {1977})}\BibitemShut {NoStop}%
\bibitem [{\citenamefont {Sandler}\ \emph {et~al.}(1994)\citenamefont
  {Sandler}, \citenamefont {Stahl}, \citenamefont {Angel}, \citenamefont
  {Lloyd-Hart},\ and\ \citenamefont {McCarthy}}]{Sandler_1994}%
  \BibitemOpen
  \bibfield  {author} {\bibinfo {author} {\bibfnamefont {D.~G.}\ \bibnamefont
  {Sandler}}, \bibinfo {author} {\bibfnamefont {S.}~\bibnamefont {Stahl}},
  \bibinfo {author} {\bibfnamefont {J.~R.~P.}\ \bibnamefont {Angel}}, \bibinfo
  {author} {\bibfnamefont {M.}~\bibnamefont {Lloyd-Hart}},\ and\ \bibinfo
  {author} {\bibfnamefont {D.}~\bibnamefont {McCarthy}},\ }\bibfield  {title}
  {\bibinfo {title} {Adaptive optics for diffraction-limited infrared imaging
  with 8-m telescopes},\ }\href {https://doi.org/10.1364/JOSAA.11.000925}
  {\bibfield  {journal} {\bibinfo  {journal} {J. Opt. Soc. Am. A}\ }\textbf
  {\bibinfo {volume} {11}},\ \bibinfo {pages} {925} (\bibinfo {year}
  {1994})}\BibitemShut {NoStop}%
\bibitem [{\citenamefont {Sharmila}\ \emph {et~al.}(2023)\citenamefont
  {Sharmila}, \citenamefont {Vermeulen},\ and\ \citenamefont
  {Datta}}]{Sharmila:2023ikb}%
  \BibitemOpen
  \bibfield  {author} {\bibinfo {author} {\bibfnamefont {B.}~\bibnamefont
  {Sharmila}}, \bibinfo {author} {\bibfnamefont {S.~M.}\ \bibnamefont
  {Vermeulen}},\ and\ \bibinfo {author} {\bibfnamefont {A.}~\bibnamefont
  {Datta}},\ }\bibfield  {title} {\bibinfo {title} {{Extracting electromagnetic
  signatures of spacetime fluctuations}},\ }\href@noop {} {\  (\bibinfo {year}
  {2023})},\ \Eprint {https://arxiv.org/abs/2306.17706} {arXiv:2306.17706
  [gr-qc]} \BibitemShut {NoStop}%
\end{thebibliography}%

\end{document}